\documentclass[journal]{IEEEtran}

\usepackage{xcolor}
\usepackage[colorlinks]{hyperref}
\usepackage[normalem]{ulem}
\usepackage{booktabs}
\usepackage[pdftex]{graphicx}
\usepackage{cuted}
\usepackage{physics}
\usepackage{cite}

\hypersetup{
	colorlinks=true, 
	linkcolor=blue!70!black, 
	citecolor=red!70!black, 
	filecolor=blue!70!black, 
	urlcolor=blue!70!black, 
}

\providecommand{\utbSmat}{\ensuremath{\M{U}}}

\definecolor{PairedA}{RGB}{166, 206, 227}
\definecolor{PairedB}{RGB}{ 31, 120, 180}
\definecolor{PairedC}{RGB}{178, 223, 138} 
\definecolor{PairedD}{RGB}{ 41, 128,  35}   
\definecolor{PairedE}{RGB}{251, 154, 153}
\definecolor{PairedF}{RGB}{182,  21,  22}   
\definecolor{PairedG}{RGB}{253, 191, 111}
\definecolor{PairedH}{RGB}{255, 127,   0}
\definecolor{PairedI}{RGB}{202, 178, 214}
\definecolor{PairedJ}{RGB}{106,  61, 154}
\definecolor{PairedK}{RGB}{255, 255, 153}
\definecolor{PairedL}{RGB}{177,  89,  40}
\providecommand{\J}{\ensuremath{\mathrm{j}}}    

\providecommand{\D}{\,\mathrm{d}}               
\providecommand{\V}[1]{\boldsymbol{#1}}         
\providecommand{\M}[1]{\mathbf{#1}}             
\providecommand{\T}[1]{\mathrm{#1}}             

\providecommand{\trans}{\mathrm{T}}

\providecommand{\UFCN}[3]{\M{u}_{#1}^{\left( #2 \right)} \left(#3\right)}

\providecommand{\Ivec}{\ensuremath{\M{I}}}

\newcommand{\ie}{\textit{i.e.}{}}
\newcommand{\eg}{\textit{e.g.}{}}

\RequirePackage[nolist]{acronym}
\newacro{MoM}[MoM]{method of moments}
\newacro{MOO}[MOO]{multiobjective optimization}
\newacro{CM}[CM]{characteristic mode}
\newacro{PEC}[PEC]{perfect electric conductor}
\newacro{PMC}[PMC]{perfect magnetic conductor}
\newacro{EP}[EP]{eigenvalue problem}
\newacro{GEP}[GEP]{generalized eigenvalue problem}
\newacro{EFIE}[EFIE]{electric field integral equation}
\newacro{SVD}[SVD]{singular value decomposition}
\newacro{RWG}[RWG]{Rao-Wilton-Glisson}
\newacro{EM}[EM]{electromagnetic}
\newacro{dof}[d-o-f]{\mbox{degrees-of-freedom}}
\newacro{MLFMA}[MLFMA]{multilevel fast multipole algorithm}
\newacro{FEM}[FEM]{finite element method}

\begin{document}

\title{Method of Moments and T-matrix Hybrid}
\author{Vit~Losenicky,
 	Lukas~Jelinek,
 	Miloslav~Capek,~\IEEEmembership{Senior~Member,~IEEE,}
 	and~Mats~Gustafsson,~\IEEEmembership{Senior~Member,~IEEE}

\thanks{Manuscript received  \today; revised \today. This work was supported by the Czech Science Foundation under project~\mbox{No.~19-06049S} and by the Grant Agency of the Czech Technical University in Prague under grant \mbox{SGS19/168/OHK3/3T/13}.}}

\markboth{Journal of \LaTeX\ Class Files,~Vol.~XX, No.~XX, \today}{Losenicky \MakeLowercase{\textit{et al.}}: Method of Moments and T-matrix Hybrid}

\maketitle

\begin{abstract}
Hybrid computational schemes combining the advantages of a method of moments formulation of a field integral equation and T-matrix method are developed in this paper. The hybrid methods are particularly efficient when describing the interaction of electrically small complex objects and electrically large objects of canonical shapes such as spherical multi-layered bodies where the T-matrix method is reduced to the Mie series making the method an interesting alternative in the design of implantable antennas or exposure evaluations. Method performance is tested on a spherical multi-layer model of the human head. Along with the hybrid method, an evaluation of the transition matrix of an arbitrarily shaped object is presented and the characteristic mode decomposition is performed, exhibiting fourfold numerical precision as compared to conventional approaches.
\end{abstract}


\IEEEpeerreviewmaketitle

\section{Introduction}
\label{Sec:Introduction}

\IEEEPARstart{N}{umerical} evaluation of \ac{EM} fields is inevitable in virtually every feasible electromagnetic design and is a driving force for the development of various computational schemes in the field of computational electromagnetism~\cite{davidson_ComputationalElectromagneticsforRFandMicrowaveEngineering}. This paper focuses on one part of this immensely vast topic, specifically on time-harmonic and full-wave descriptions of open multi-scale problems where several objects of varying electrical size interact in otherwise open space.

One of the basic, full-wave numerical techniques for solving open boundary electromagnetic problems is the \ac{MoM} formulation of field integral equations~\cite{Harrington_FieldComputationByMoM}. This method is most popular in its surface version~\cite{RaoWiltonGlisson_ElectromagneticScatteringBySurfacesOfArbitraryShape} which assumes highly conducting bodies or surface equivalence treatment~\cite{Harrington_TimeHarmonicElmagField}. In these cases, the method reveals its greatest advantages: accuracy, low number of unknowns and computational efficiency all of which are associated with the discretization of surface current densities. The major weakness of this method is the use of a fully populated system matrix which leads to an undesirable increase in memory and computational requirements with increasing complexity and electrical size of the problem\cite{Kolundzija_ElectromagneticModelingofCompositeMetallicandDielectricStructures}.

The aforementioned difficulties are most commonly mitigated via the~\ac{MLFMA}~\cite{ErgulGurel_MLFMA}, the adaptive cross approximation algorithm (ACAA)~\cite{Zhao_etal_2005_ACA}, and the characteristic basis functions method (CBFM)~\cite{PrakashMittra_CHBFM}. The~\ac{MLFMA} mitigates memory and computational requirements while maintaining accuracy using interaction via multi-pole expansion. The ACAA reduces the computational complexity and memory requirements like the~\ac{MLFMA} while remaining independent of the integration kernel. The CBFM deals with the weaknesses from a different prospective and uses high-level basis functions defined on macro domains leading to matrix size reduction. An alternative is to implement piecewise-defined high-order basis functions~\cite{Graglia_HigherOrderInterpolatoryVectorBasesforComputationalElectromagnetics} which can significantly reduce the number of unknowns~\cite{Kolundzija_ElectromagneticModelingofCompositeMetallicandDielectricStructures}.

The description via volume integral equations can also be used in cases of complex material distributions, but, with the exception of bodies of revolution\cite{Mautz_RadiationandScatteringFromBodiesofRevolution} and translation symmetries~\cite{Wandzura_FastFourierTransformTechniquesforSolvingtheEFIEforPeriodicBody, Polimeridis+etal2014}, the memory and computational requirements forbid its use in realistic scenarios. In such cases, the \ac{FEM}~\cite{Volakis_1998_FEM} operating over differential equations is typically used instead. Unfortunately, there is no universal method, and the \ac{FEM} has its problems, the major ones being the accurate treatment of open boundaries and electrically large objects~\cite{Volakis_1998_FEM}.

When multi-scale objects interact, the leading approach is the use of hybrid methods attempting to use particular evaluation schemes only in situations when they are effective. In the case of open electromagnetic problems with only medium electric size, the hybrid combining the \ac{FEM} for regions of high material complexity and the integral equation method for treating highly conducting objects and open boundaries is common \cite{Jin_AHybridFinitEelementMethodforScatteringandRadiation, Gedney_ACombinedFEMMOMApproachtoAnalyzethePlaneWaveDiffraction, Ali_AHybridFEMMOMTechniquefroElectromagneticScatteringandRadiation, Chio_Hybrid3DFEMOMAnalysusforAntennaswithThinDielectricCover, Liu_ANovelHybridizationofHighOrderFEandBEM} and is a part of several commercial~\cite{feko2021, HFSS} as well as in-house\cite{Yun_EMCApplicationsoftheEMAP5HybridFEMMOMCode} implementations.

Despite the great versatility of the aforementioned hybrid method, the description of the interaction of electrically small and electrically large material objects is nevertheless problematic. However, in many cases, the electrically large object can be approximated by a spheroidal geometry, a situation when the T-matrix method~\cite{Waterman1965} can be advantageously used. The most important property in this respect is the use of spherical vector waves as entire domain basis functions. This allows for the compressed description of spheroidal-like bodies~\cite{LiKangLeong_SpheroidalWaveFunctionsInEMTheory, Stratton_SpheroidalWaveFunctions} or even a description by a diagonal matrix in the case of a spherical multi-layer, which is a common approximation used in electromagnetism.

The purpose of this paper is to approach the multi-scale problem by combining the \ac{MoM} formulation of the electric field integral equation and the T-matrix method in which the T-matrix method is used to describe interacting electrically large objects of simple shape efficiently. 
The coupling between the two methods is described by spherical vector waves~\cite{Hansen_SphericalNearFieldAntennaMeasurements}. The major advantage of the proposed computational scheme is its ability to provide a full system matrix of the electrically small object in the otherwise multi-scale scenario.

The paper is organized as follows. The \ac{MoM} formulation of the electric field integral equation and the T-matrix method are briefly recapitulated in Sections~\ref{subsec:MoM} and~\ref{subsec:Tmatrix}, respectively, and the ``external'' and ``internal'' formulations of their hybrid method are afterwards developed in Sections~\ref{sec:HybridExt} and~\ref{sec:HybridInt}. Section~\ref{Sec:ResultsI} provides the numerical verification of the hybrid method. The unification of ``external'' and ``internal'' formulation is presented in Section~\ref{sec:hybridUnified} and is later verified in Section~\ref{Sec:ResultsII}. Various aspects of the hybrid method are discussed in Section~\ref{sec:Discussion} and the paper is concluded with Section~\ref{Sec:Conclusion}.

\section{Basic Constituents of the Hybrid Method}
\label{sec:hybridConstituents}

This section briefly reviews the \ac{MoM} formulation of field integral equations~\cite{Harrington_FieldComputationByMoM} and the T-matrix method~\cite{Waterman1965}, the basic constituents of which are subsequently hybridized. 

\subsection{Method of Moments Formulation to Electric Field Integral Equation}
\label{subsec:MoM}
\begin{figure}[t]
    \centering
    \includegraphics[height=2.571cm]{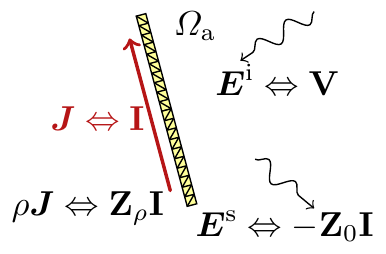}
    \caption{An illustration of the \ac{MoM} formulation for the electric field integral equation where the underlying scatterer is discretized into a set of elementary cells. The figure also interrelates field variables~$\V{J}$, $\rho\V{J}$, $\V{E}^\T{s}$, and~$\V{E}^\T{i}$ with their discretized counter parts~$\M{I},\M{Z}_\rho\Ivec$, $-\M{Z}_0 \Ivec$, and~$\M{V}$, respectively.}
    \label{fig:MoMIllustration}
\end{figure}

The electric field integral equation\footnote{Time-harmonic steady state of convention~$\T{exp}\{\T{j}\omega t\}$, where~$\omega$ is the angular frequency and~$\T{j}$ the imaginary unit, is assumed throughout the paper.} is formed by relating the polarization (or conduction) current density in a material object to the total electric field as
\begin{equation}
    \label{eq:EFIE}
    \V{E}^\T{s}\left(\V{J}\right) + \V{E}^\T{i} = \rho \V{J},
\end{equation}
with 
\begin{equation}
    \label{eq:Es}
   \V{E}^\T{s}\left(\V{J}\right) = - \T{j}kZ \left \langle \V{G}_\T{e}, \V{J} \right \rangle
\end{equation}
being the scattered field produced by current distribution~$\V{J}$, $\V{E}^\T{i}$ being the incident field, $k$ being the background wavenumber\footnote{Any homogeneous lossless dielectric material can be considered as background in this formulation.}, $Z$ being the background wave impedance, $\V{G}_\T{e}$ being the dyadic Green's function for electric fields, and~$\rho \left( \V{r} \right)$ being the complex resistivity of the underlying material. The relation~\eqref{eq:Es} also utilizes a symmetric product (reaction)~\cite{Rumsey_ReactionConceptInElectromagneticTheory}, which is, for volume distributions, defined as
\begin{equation}
    \label{eq:ipdS}
    \left \langle \V{A},\V{B} \right \rangle \equiv \int_{V}\V{A}\cdot\V{B}\D{V}.
\end{equation}
with $V$ indicating the support of fields~$\V{A}$ and $\V{B}$.
An analogous symmetry product can be defined for surface current distribution by integration over the corresponding surface.
A similar distinction should be made over complex resistivity~$\rho$, which typically refers to a volumetric case, while for surface current distribution is the resistivity commonly substituted by a surface impedance~\cite{Jackson_ClassicalElectrodynamics,SenoirVolakis_ApproximativeBoundaryConditionsInEM}.

To solve~\eqref{eq:EFIE} numerically, the original material object is decomposed into a set of elementary cells (\eg{}, triangles, tetrahedrons), see~Fig.~\ref{fig:MoMIllustration}, together with the current density which is rewritten as a weighted sum of basis functions~$\{\V{\psi}_n\}$
\begin{equation}
    \label{eq:Jexp}
    \V{J}\left(\V{r}\right)\approx\sum_{n=1}^N I_n\V{\psi}_n\left(\V{r}\right).
\end{equation}
Substituting~\eqref{eq:Jexp} into~\eqref{eq:EFIE} and employing the Galerkin testing technique~\cite{ChewTongHu_IntegralEquationMethodsForElectromagneticAndElasticWaves} results in a linear system of equations 
\begin{equation}
    \label{eq:MoMip}
    -\Big [ \big\langle \V{\psi}_m, \V{E}^\T{s}\left(\V{\psi}_n\right) \big\rangle \Big]\M{I} + \Big [ \big\langle \V{\psi}_m, \rho \V{\psi}_n\big\rangle \Big]\M{I} = \Big [ \big\langle \V{\psi}_m, \V{E}^\T{i}\big\rangle \Big],
\end{equation}
which can be rewritten as
\begin{equation}
    \label{eq:MoM}
    \left( \M{Z}_0 + \M{Z}_\rho \right) \M{I} = \M{V},
\end{equation}
with $\M{Z}_0$ being the impedance matrix of the radiation part of the system, $\M{Z}_\rho$ being the impedance matrix of the material part of the system, $\M{V}$ being the excitation vector, and $\M{I}$ being the unknown vector of current expansion coefficients.

Field integral equations are part of many commercial~\cite{feko2021, cst2021, wipld2020, cemone2020} and academic~\cite{atom} electromagnetic simulators with the major advantage being the implicit incorporation of boundary conditions into Green's function~\cite{Dudley_MathematicalFoundationsForElectromagneticTheory}, which renders the \ac{MoM} formulation of field integral equations as an excellent method for open problems. The resulting operator matrices are dense but typically much smaller as compared to, \eg{}, \ac{FEM}, yet fully describe the \ac{EM} properties of the radiator. This allows eigenvalue problems to be formulated, such as characteristic mode decomposition~\cite{HarringtonMautz_TheoryOfCharacteristicModesForConductingBodies} or the optimal current densities representing fundamental bounds on \ac{EM} metrics~\cite{GustafssonTayliEhrenborgEtAl_AntennaCurrentOptimizationUsingMatlabAndCVX} to be found. The disadvantage is an undesired sextic growth of memory requirements with the electrical size of a radiator. With radiation being the dominating interaction, memory requirement can be reduced and a solution can be accelerated via the~\ac{MLFMA}~\cite{ErgulGurel_MLFMA} or ACAA~\cite{Zhao_etal_2005_ACA}, which lead to linearithmic time and cubic storage, but forbids the direct evaluation of the fundamental bounds. Instead, the individual obstacles can be decoupled and treated independently, for example, by combining \ac{MoM} with piecewise defined basis functions~\cite{PetersonRayMittra_ComputationalMethodsForElectromagnetics} and the T-matrix method for spherical waves~\cite{Kristensson_ScatteringBook} into one system of equations which are still compatible with mode decomposition techniques, convex optimization, and other methods explicitly requiring the impedance matrix. 

\subsection{T-matrix Method}
\label{subsec:Tmatrix}
\begin{figure}[t]
    \centering
    \includegraphics[height=4.174cm]{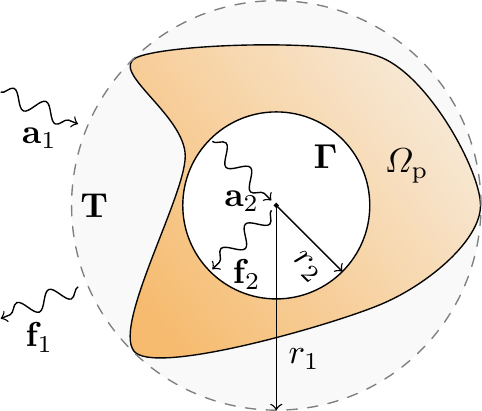}
    \caption{Scattering properties of object $\varOmega_\T{p}$ are described by scattering operator~$\M{T}$ for external problems and by operator~$\M{\Gamma}$ for internal problems. Since both operators are based on the expansion of the dyadic Green's function into spherical vector waves, all field quantities are valid only in the region $r<r_1$ and $r>r_2$.}
    \label{fig:TGIllustration}
\end{figure}

The T-matrix method is used to solve a similar scenario as in~\eqref{eq:EFIE} using spherical harmonics as entire-domain basis functions making it extremely effective in solving scattering from spheroidal particles~\cite{MishchenkoTravis_TMatrixComputationsofLightScatteringbyNonsphericalParticlesReview}. The T-matrix method~\cite{Waterman1965} starts with an expansion of the incident and scattered electric field external to the material object, see Fig.~\ref{fig:TGIllustration}, into a set of spherical vector waves as~\cite{Hansen_SphericalNearFieldAntennaMeasurements}
\begin{equation}
    \label{eq:EiExp1}
    \V{E}^\T{i}\left(\V{r}\right) = k\sqrt{Z}\sum_\alpha a_{1,\alpha} \,\UFCN{\alpha}{1}{k\V{r}},
\end{equation}
\begin{equation}
    \label{eq:EsExp1}
    \V{E}^\T{s}\left(\V{r}\right) = k\sqrt{Z}\sum_\alpha f_{1,\alpha}\,\UFCN{\alpha}{4}{k\V{r}},
\end{equation}
where $\M{a}_1$ is a vector of expansion coefficients of the incident field into regular spherical vector waves, $\M{f}_1$ is a vector of expansion coefficients of the scattered field into a set of out-going spherical vector waves, and~$\UFCN{\alpha}{p}{k\V{r}}$ are the spherical vector waves defined in Appendix~\ref{App:sphericalExpansion}. Note that such an expansion is only valid outside a sphere circumscribing the material object, see~Fig.~\ref{fig:TGIllustration}. Since a homogeneous background material is assumed outside the circumscribing sphere, it is possible to define transition matrix~$\M{T}$ via
\begin{equation}
    \label{eq:Tmethod}
    \M{f}_1 = \M{T}\M{a}_1,
\end{equation}
forming a linear system analogous to~\eqref{eq:MoM}. For spherical objects, matrix~$\M{T}$ can be obtained analytically, see~Appendix~\ref{App:TmatSphere}. For material objects of a general shape matrix~$\M{T}$ can be evaluated using the Null-field method~\cite{Waterman1965,Mishchenko_ScatteringAbsorptionandEmissionogLightbySmallParticles} or using matrix~$\M{Z}$ as described later, see~Appendix~\ref{App:ZtoT} for final formulas.

The internal scattering problem defined via the expansion coefficients~$\M{a}_2$ and~$\M{f}_2$, see~Fig.~\ref{fig:TGIllustration}, can also be solved by this methodology with the only change being that the incident field is formed by spherical waves outgoing from the origin while the scattered field is represented by regular spherical waves. With this change, the internal scattering problem can be described via
\begin{equation}
    \label{eq:Gammamethod}
    \M{a}_2 = \M{\Gamma}\M{f}_2.
\end{equation}

A major distinction between matrices~$\M{T}$ or $\M{\Gamma}$ and matrix~$\M{Z}$ is the fact that neither~$\M{T}$ nor $\M{\Gamma}$ are invertible and that a distinction must be made between external and internal problems.

The description via matrix~$\M{T}$ is widely used in the field of electromagnetic scattering~\cite{MishchenkoTravis_TMatrixComputationsofLightScatteringbyNonsphericalParticlesReview} and, unlike the impedance matrix from the previous section, matrix~$\M{T}$ only accounts for the scattering reaction on the incident field produced externally to the scatterer. This method is especially efficient for scatterers of spheroidal shape when the system matrix can be evaluated analytically~\cite{Kristensson_ScatteringBook}. On the other hand, this simplification prohibits near fields within the sphere circumscribing the scatterer to be studied or localized feeding ports on antennas to be defined since no sources can be present in region~$\varOmega_\T{p}$. It is also not known how to evaluate fundamental bounds in such a description as there is no direct access to the internal degrees of freedom describing contrast current density within the scatterer.

\section{Hybrid Method}
\label{sec:Hybrid}
To characterize scattering from a complex-shaped object of electrical size not exceeding a few wavelengths or containing discrete feeding ports, the impedance matrix based on triangular or tetrahedral elements is the tool of choice mostly for numerical stability of this scheme regardless of the shape's complexity. On the contrary, when dealing with an electrically large regular-shaped and passive scatterer, the description via matrix~$\M{T}$ offers many advantages, such as a great model order reduction due to the use of appropriate entire domain basis functions. The purpose of this section is to combine the strengths of both methods. The resulting technique is shown to be especially fast and flexible for problems where canonical models of obstacles (described by matrix~$\M{T}$) are sufficient to grasp the most important interactions with radiator described by matrix~$\M{Z}$ such as those appearing in the study of implantable antennas. The resulting formulation still allows for modal decomposition or the evaluation of fundamental bounds via the convex optimization of current density~\cite{GustafssonTayliEhrenborgEtAl_AntennaCurrentOptimizationUsingMatlabAndCVX}.

\subsection{External Formulation}
\label{sec:HybridExt}
\begin{figure}[t]
    \centering
    \includegraphics[height=4.174cm]{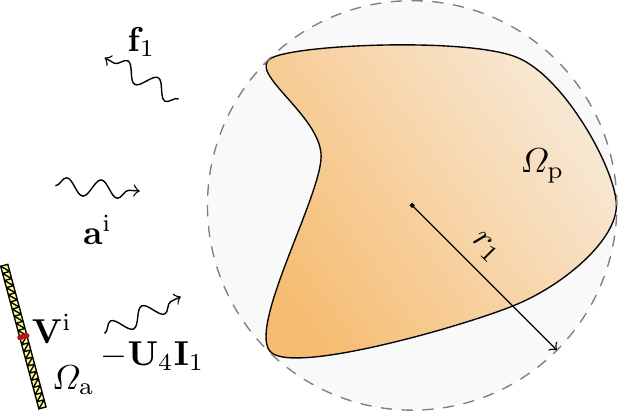}
    \caption{An illustration of the hybrid method for the exterior of object~$\varOmega_\T{p}$. The interaction between objects~$\varOmega_\T{p}$ and~$\varOmega_\T{a}$ is realized via spherical vector waves weighted by coefficients collected in vectors $\M{f}_{1,2}$ and $\M{a}_{1,2}$. Note that the complete vector~$\M{a}_1$ that excites object~$\varOmega_\T{p}$ is given by~\eqref{eq:arad}, while the complete vector~$\M{V}$ exciting object~$\varOmega_\T{a}$ is given by~\eqref{eq:Vf}.}
    \label{fig:external}
\end{figure}

Let us assume two objects~$\varOmega_\T{p}$ and $\varOmega_\T{a}$ depicted in Fig.~\ref{fig:external}. Object~$\varOmega_\T{p}$ is situated around the origin of the coordinate system and represented by matrix~$\M{T}$. Object~$\varOmega_\T{a}$ is characterized by the matrix~$\M{Z} = \M{Z}_0 + \M{Z}_\rho$ and placed so that it does not intersect the sphere circumscribing object~$\varOmega_\T{p}$. Assume further an impressed excitation~$\V{V}^\T{i}$ (impressed electric field on object~$\varOmega_\T{a}$ via, \eg{}, a delta gap source) and~$\M{a}^\T{i}$ (a set of impinging spherical waves).

To describe the problem in full, it is convenient to form a complete vector~$\M{V}$ exciting object~$\varOmega_\T{a}$ from impressed field~$\M{V}^\T{i}$, impressed field~$\M{a}^\T{i}$, and field vector~$\M{f}_1$ produced by object~$\varOmega_\T{p}$. Analogously, it is convenient to form a complete vector~$\M{a}_1$ from impressed field~$\M{a}^\T{i}$ and from spherical waves~$\M{a}$ produced by object~$\varOmega_\T{a}$.

Let us first focus on the complete vector~$\M{V}$. The field produced by object~$\varOmega_\T{p}$ is given by~\eqref{eq:EsExp1}. The impressed field~$\V{a}^\T{i}$ is formed by a relation equivalent to~\eqref{eq:EiExp1}. Adding these two fields, substituting into the right-hand side of~\eqref{eq:MoMip} and subsequently adding~$\M{V}^\T{i}$ leads to 
\begin{equation}
    \label{eq:Vf}
    \M{V} = \M{V}^\T{i} + 
    \utbSmat_1^\T{T}\M{a}^\T{i} + 
    \utbSmat_4^\T{T}\M{f}_1,
\end{equation}
where matrices $\utbSmat_p$, $p\in \left\{1,\dots, 4\right\}$, defined as
\begin{equation}
    \label{eq:Smatrix}
    \utbSmat_p = k\sqrt{Z} \left[ \left \langle\M{u}_\alpha^{\left(p\right)}, \V{\psi}_n \right \rangle \right],
\end{equation}
project spherical waves onto basis functions~$\V{\psi}_n$, see Appendix~\ref{App:sphericalExpansion} for details. Notice that matrices analogous to~\eqref{eq:Smatrix} are used in the null field method~\cite{Kristensson_ScatteringBook} and\footnote{Note the use of normalization of spherical waves used in this paper.} in\cite[(10-11)]{2013_Kim_TAP} \cite[(11)]{2017_Markkanen_JQSRT}. 

The second relation is obtained expanding the field~\eqref{eq:Es} produced by object~$\varOmega_\T{a}$ into spherical expansion~\eqref{eq:EiExp1} and adding an impressed field~$\M{a}^\T{i}$, which results in
\begin{equation}
    \label{eq:arad}
    \M{a}_1 = \M{a}^\T{i} -\utbSmat_4\M{I}.
\end{equation}
Notice that the relation~\eqref{eq:arad} uses the spherical expansion of the dyadic Green's function into spherical vector waves, see~Appendix~\ref{App:sphericalExpansion}. Combining \eqref{eq:MoM}, \eqref{eq:Tmethod}, \eqref{eq:Vf} and \eqref{eq:arad} together, an equation system
\begin{equation}
\label{eq:ExternalFull}
    \mqty[\M{Z} & -\utbSmat_4^\T{T} & \M{0} \\ -\utbSmat_4 & \M{0} & -\M{1} \\
    \M{0} & -\M{1} & \M{T} ]
\mqty[ \M{I} \\ \M{f}_1 \\ \M{a}_1
        ] =
    \mqty[\M{V}^\T{i} + \utbSmat_1^\T{T} \M{a}^\T{i} \\ -\M{a}^\T{i} \\\M{0} ]
\end{equation}
is formed providing a direct solution to all unknown quantities. 

It might be advantageous to partially resolve the system~\eqref{eq:ExternalFull} for a particular unknown variable in many cases. A typical scenario might be object~$\varOmega_\T{a}$ being studied in the presence of parasitic scatterer~$\varOmega_\T{p}$. Eliminating unknowns~$\M{a}_1$ and $\M{f}_1$ from~\eqref{eq:ExternalFull} then gives
\begin{equation}
\label{eq:ExternalI}
    \left(\M{Z}+\utbSmat_4^\T{T}\M{T}\utbSmat_4\right)\M{I}= \M{V}^\T{i} +  \left(\utbSmat_1^\T{T} +  \utbSmat_4^\T{T}\M{T}\right)\M{a}^\T{i},
\end{equation}
where the only unknown is current~$\M{I}$ on object~$\varOmega_\T{a}$. Notice that this last equation is equivalent to applying the \ac{MoM} to the electric field integral equation for scatterer~$\varOmega_\T{a}$ using a Green's function accounting for object~$\varOmega_\T{p}$. This last formulation is also prepared for the development of fundamental bounds based on a controllable current~\cite{Gustafsson_OptimalAntennaCurrentsForQsuperdirectivityAndRP} supported in region~$\varOmega_\T{a}$ in the presence of an uncontrollable scatterer residing in region~$\varOmega_\T{p}$.

\subsection{Internal Formulation}
\label{sec:HybridInt}
\begin{figure}[t]
    \centering
    \includegraphics[height=3.771cm]{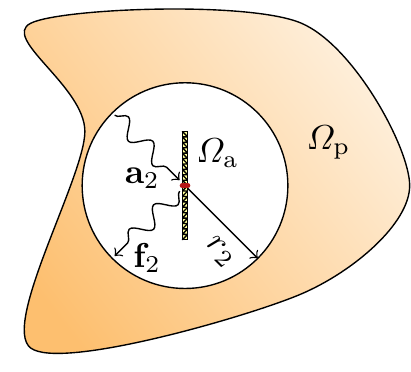}
    \caption{An internal definition of the hybrid method involves an electromagnetic radiator (object~$\varOmega_\T{a}$) inserted into the cavity of radius $r_2$ immersed in object~$\varOmega_\T{p}$. The interaction between the radiator and the scatterer is provided by two vectors that stand for the weighting coefficients of regular~$\M{a}_2$ and out-going~$\M{f}_2$ spherical vector waves.}
    \label{fig:internal}
\end{figure}

A dual scenario, depicted in~Fig.~\ref{fig:internal}, where an impressed spherical excitation has been omitted for clarity, can be solved analogously.
For this arrangement, object~$\varOmega_\T{a}$ is centered at the origin of the coordinate system, while object~$\varOmega_\T{p}$ contains a cavity surrounding it meeting the requirement that a circumscribing sphere of object~$\varOmega_\T{a}$ fits inside the cavity. The total excitation of object~$\varOmega_\T{p}$ reads
\begin{equation}
\label{eq:f2}
    \M{f}_2 = - \utbSmat_1\M{I},
\end{equation}
while the total excitation of object~$\varOmega_\T{a}$ reads
\begin{equation}
    \M{V} = \M{V}^\T{i} + \utbSmat_1^\T{T} \M{a}_2.
\end{equation}
The algebraic expression for the problem defined in this way can therefore be written as
\begin{equation}
\label{eq:InternalFull}
    \mqty[\M{Z} & -\utbSmat_1^\T{T} & \M{0} \\
            -\utbSmat_1 & \M{0} & -\M{1} \\
            \M{0} & -\M{1} & \M{\Gamma} ]
\mqty[ \M{I} \\ \M{a}_2 \\ \M{f}_2
        ] =
    \mqty[ \M{V}^\T{i} \\ \M{0} \\ \M{0} ].
\end{equation}

An equation analogous to~\eqref{eq:ExternalI}, with the sole unknown being current~$\M{I}$, is given by
\begin{equation}
\label{eq:InternalI}
    \left(\M{Z}+\utbSmat_1^\T{T}\M{\Gamma}\utbSmat_1\right)\M{I}= \M{V}^\T{i}.
\end{equation}

It is important to note that even though the internal scattering operator $\M{\Gamma}$ has a lot of common features with the transition matrix, it is not possible to find matrix~$\M{\Gamma}$ simply by the inversion of the transition matrix. 

\section{Results: Numerical Validation}
\label{Sec:ResultsI}
In this section, an evaluation of the transition matrix based on the impedance matrix, as well as the hybrid method combining \ac{MoM} and the T-matrix method, are used for the analysis of several canonical problems. The results are compared with previously published results~\cite{TayliEtAl_AccurateAndEfficientEvaluationofCMs} or with the outcomes of the commercially available electromagnetic solvers FEKO~\cite{feko2021} and CST~\cite{cst2021}.

\subsection{Characteristic Modes}
Evaluation of matrix~$\M{T}$ is validated first. The test is based on the comparison of eigenvalues of matrix~$\M{T}$ and characteristic numbers obtained by characteristic mode decomposition~\cite{HarringtonMautz_TheoryOfCharacteristicModesForConductingBodies} for a~\ac{PEC} shell for which characteristic numbers are known analytically~\cite{CapekEtAl_ValidatingCMsolvers}. Matrix~$\M{T}$ is known analytically as well in this case, see~\eqref{eq:TPEC}. Nevertheless, in order to prove the validity of the theory presented in the previous section,
matrix~$\M{T}$ is evaluated numerically according to Appendix~\ref{App:ZtoT}.

The characteristic mode decomposition~\cite{HarringtonMautz_TheoryOfCharacteristicModesForConductingBodies,HarringtonMautzChang_CharacteristicModesForDielectricAndMagneticBodies} is for a lossless scatterer\footnote{For a lossy scatterer, the right-hand side of~\eqref{eq:CM1} might be changed to~$\M{R}_0$, but in such a case the relation to eigensolutions of matrix~$\M{T}$ is lost.} defined by the generalized eigenvalue problem 
\begin{equation}
    \label{eq:CM1}
    \M{X}\M{I}_n = \lambda_n\M{R}\M{I}_n
\end{equation}
in which matrices $\M{R}=\T{Re}\{\M{Z}\}$ and $\M{X}=\T{Im}\{\M{Z}\}$ are the real and imaginary parts of the impedance matrix respectively, $\lambda_n$~is the characteristic number, and $\M{I}_n$ is a characteristic mode. The relation between the characteristic numbers and eigenvalues of matrix~$\M{T}$ can be expressed as 
\begin{equation}
\label{eq:xiLambdaRel}
    t_n =-\frac{1}{1+\T{j}\lambda_n}
\end{equation}
with $t_n$ being the eigenvalues of matrix~$\M{T}$. The derivation of relation~\eqref{eq:xiLambdaRel} is detailed in Appendix~\ref{app:CM}. Notice that the identification of characteristic numbers with eigenvalues of matrix~$\M{T}$ has also been provided in~\cite[(18)]{Garbacz_TCMdissertation}.

\begin{figure}[t]
    \centering
    \includegraphics[width=8.5cm]{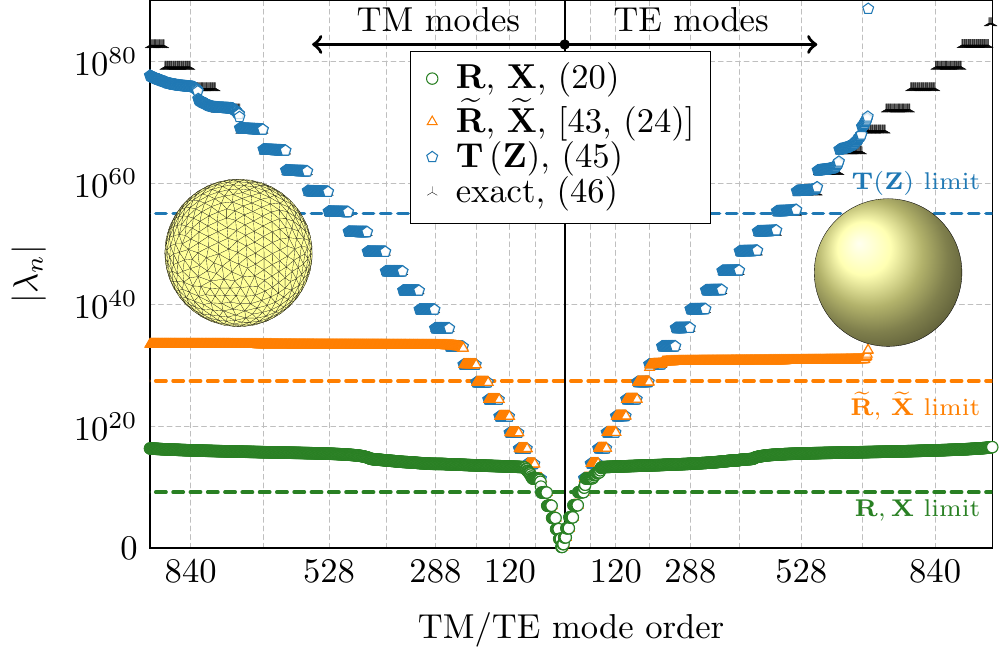}
    \caption{Magnitudes of the characteristic numbers of a spherical shell ($ka=0.5$). Presented results were obtained using four different methods: the classic procedure~\eqref{eq:CM1} depicted by green circles, the method presented in~\cite{TayliEtAl_AccurateAndEfficientEvaluationofCMs} depicted by orange triangles, the method presented in this paper~\eqref{eq:CM5} with transformation by~\eqref{eq:xiLambdaRel} depicted by blue circles, and the analytic prescription~\cite{CapekEtAl_ValidatingCMsolvers} depicted by black stars.}
    \label{fig:Sphere_T_CM}
\end{figure}

The result of this first test is depicted in Fig.~\ref{fig:Sphere_T_CM} for a spherical shell with electrical size ~$ka=0.5$, $a$ being the radius of the sphere. The spherical shell is discretized into $1376$~triangles and~\ac{RWG} basis functions~\cite{RaoWiltonGlisson_ElectromagneticScatteringBySurfacesOfArbitraryShape} from~\eqref{eq:Jexp} are used as a basis~$\left\{\V{\psi}_n\left(\V{r}\right)\right\}$. The characteristic numbers are computed in three different ways. First is the classical procedure of a generalized eigenvalue problem~\eqref{eq:CM1}. Second is the improved approach using a modified impedance matrix and singular value decomposition~\cite{TayliEtAl_AccurateAndEfficientEvaluationofCMs} to enhance numerical precision. Finally, characteristic numbers are evaluated using an eigenvalue decomposition of matrix~$\M{T}$~\eqref{eq:CM5} and its relation to characteristic numbers~\eqref{eq:xiLambdaRel}. Analytically known characteristic numbers~\eqref{eq:CM6} further supplement these three numerical solutions.

The results depicted in Fig.~\ref{fig:Sphere_T_CM} not only prove the validity of the matrix~$\M{T}$ evaluation but also show a superior accuracy of the third method, \ie{}, of evaluating characteristic numbers from eigenvalues of matrix~$\M{T}$. The precision enhancement is immense, reaching a dynamic range of~$10^{70}$ and doubling the dynamic range of the second-best numerical procedure.

\begin{figure}[t]
    \centering
    \includegraphics[width=8.5cm]{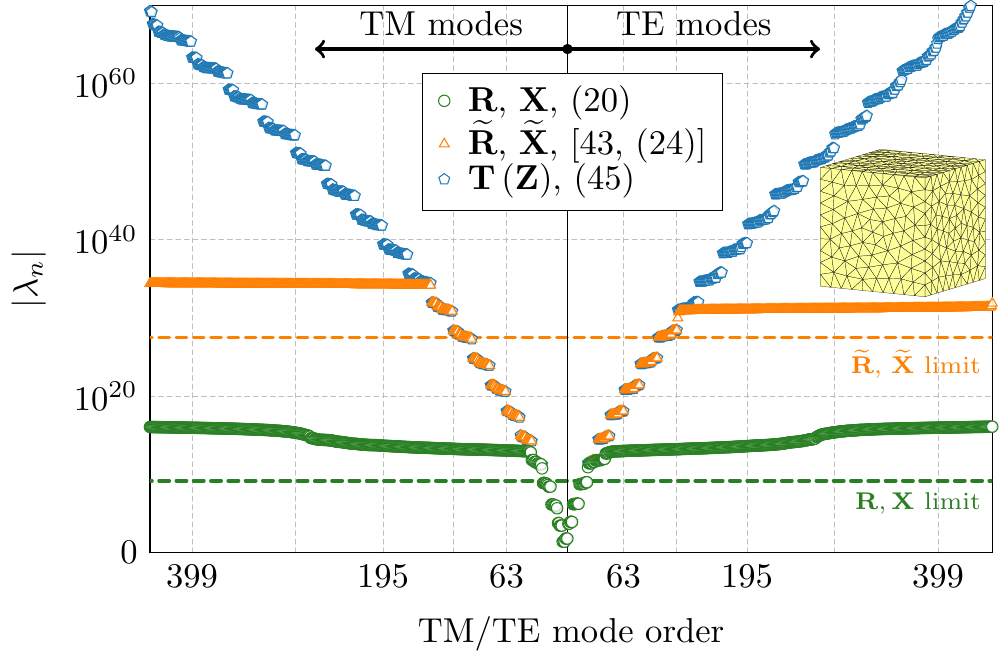}
    \caption{Characteristic numbers of a~\ac{PEC} cube ($ka=0.5$). The data traces are the same as in Fig.~\ref{fig:Sphere_T_CM}.}
    \label{fig:Cube_T_CM}
\end{figure}

The second numerical test uses the evaluation of characteristic numbers of a~\ac{PEC} cube discretized into~$912$ triangles and circumscribed by a sphere of radius~$a$. In this case, characteristic numbers are not known analytically. The evaluation of matrix~$\M{T}$ is based on Appendix~\ref{App:ZtoT}.
The results obtained from the second test are depicted in Fig.~\ref{fig:Cube_T_CM}. The numerical precision of characteristic numbers evaluated via~\eqref{eq:xiLambdaRel} is once more superior and does not saturate as with the other methods. 

\subsection{Hybrid Method}
\label{Sec:ResultsI:HybridMethod}
This section aims to verify the hybrid method described in Section~\ref{sec:Hybrid}. The two scenarios, ``external'' in Fig.~\ref{fig:external} and ``internal'' in Fig.~\ref{fig:internal}, are tested separately.

The ``external'' case describing the relation between a passive body and an active radiator in its exterior is validated first. The test is built on the analysis of the interaction between a spherical shell and a dipole antenna. The spherical shell is centered at the origin, has outer radius~$r$, thickness~$d$, and is made of a material with relative permittivity~$\varepsilon_\T{r}$. At distance~$g$ from the surface of the shell, a strip dipole antenna of length~$\ell$ and width~$w$ is located. For the sake of simplicity, the dipole antenna is modeled as a PEC strip fed in the middle by a delta-gap feed.

\begin{figure}[t]
    \centering
    \includegraphics[width=8.5cm]{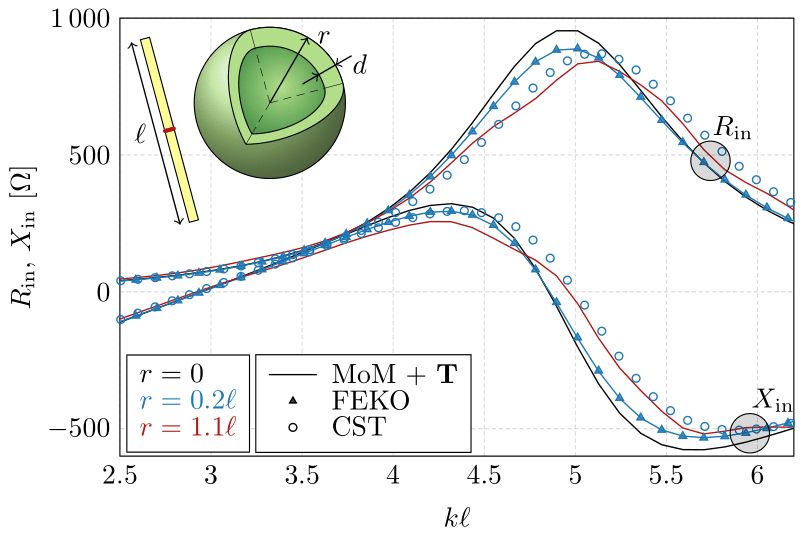}
    \caption{Input impedance~$Z_\T{in} = R_\T{in} + \T{j} X_\T{in}$ of a dipole with length~$\ell$ and width~$w=\ell/50$ discretized into 396 triangles. The geometric center of the dipole is located at distance~$g=\ell/20$ from the surface of a spherical shell with outer radius $r$ and thickness $d/\ell=0.05$ made of a material with relative permittivity~$\varepsilon_\T{r}=5-0.5\J$. The black line represents input impedance for the dipole in vacuum. The blue line corresponds to a spherical shell with outer radius~$r=0.2\ell$ and evaluation via~\eqref{eq:ExternalFull}. This result is compared with the solution supplied by the combination of \ac{MoM} and \ac{FEM} solver of FEKO~\cite{feko2021} and time-domain solver of CST~\cite{cst2021}. Additionally, a result of~\eqref{eq:ExternalFull} for the spherical shell with radius~$r=1.1\ell$ is shown (red lines). }
    \label{fig:T_out_inputImpedance}
\end{figure}

\begin{figure}[t]
    \centering
    \includegraphics[width=8.5cm]{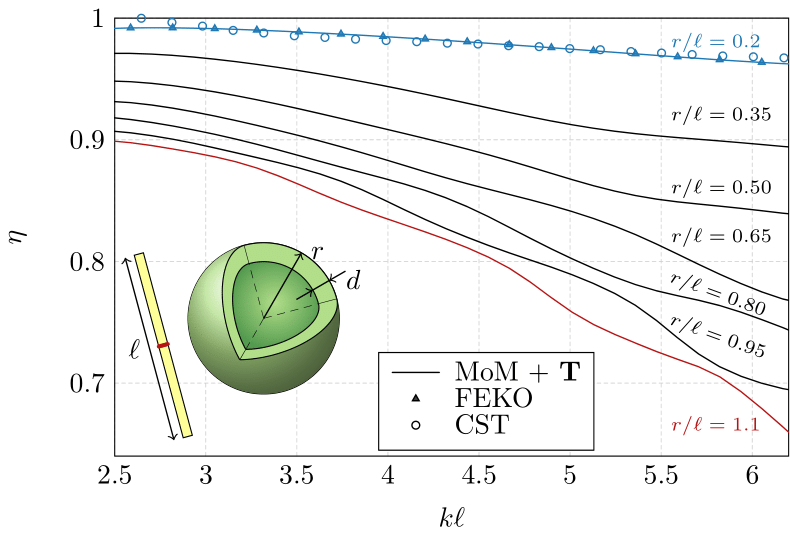}
    \caption{The radiation efficiency of the dipole in the vicinity of the spherical shell of varying radius~$r$. The setup is identical to Fig.~\ref{fig:T_out_inputImpedance}. In the case of a spherical shell of outer radius~$r=0.2\ell$, the result is compared with the FEKO and CST solver. Red and blue lines are the same as in Fig.~\ref{fig:T_out_inputImpedance}.}
    \label{fig:T_out_efficiency}
\end{figure}

The verification is based on two quantities relevant for antenna performance: input impedance and radiation efficiency. The results depicted in~Figs.~\ref{fig:T_out_inputImpedance} and~\ref{fig:T_out_efficiency} verify the method by comparing results of~\eqref{eq:ExternalFull} with results obtained by the combination of \ac{MoM} and \ac{FEM} in FEKO~\cite{feko2021} and the time-domain solver of CST~\cite{cst2021}. Both figures suggest perfect agreement with the FEKO solver. A small deviation in the case of the input impedance evaluated by the CST solver can be attributed to a different model of the feed which consists of a physical gap in the CST solver. More details about solver settings can be found in Appendix~\ref{app:Solvers}.
The part of the hybrid method corresponding to the T-matrix method is based on the Mie series solution, see~\cite{Yang_ImprovedRecursiveAlgorithmforLightScatteringbyaMultilayeredSphere} and the references therein, and it is, hence, numerically inexpensive to modify the size of the lossy sphere as depicted in Fig.~\ref{fig:T_out_efficiency}.

Figure~\ref{fig:T_out_inputImpedance} indicates that the input impedance in the vicinity of the resonance of the dipole is relatively stable and it is not significantly affected by its environment. For this reason, in addition to input impedance, radiation efficiency was also evaluated\footnote{See~Appendix~\ref{app:PowerBalance} for details on power balance in the hybrid scenario described in this paper.}. Since the dipole is lossless, any deviation of radiation efficiency from unity is induced by the spherical shell. The results are depicted in Fig.~\ref{fig:T_out_efficiency} where perfect agreement between the solution via~\eqref{eq:ExternalFull} and via the \ac{FEM} solver of FEKO can once more be seen. Unlike input impedance, which does not depend much on the radius of the absorbing shell, the radiation efficiency of the system decreases significantly with the increasing ratio of~$r/l$. This is caused by the increasing efficiency of thicker shells to absorb radiation.

\begin{figure}[t]
    \centering
    \includegraphics[width=8.5cm]{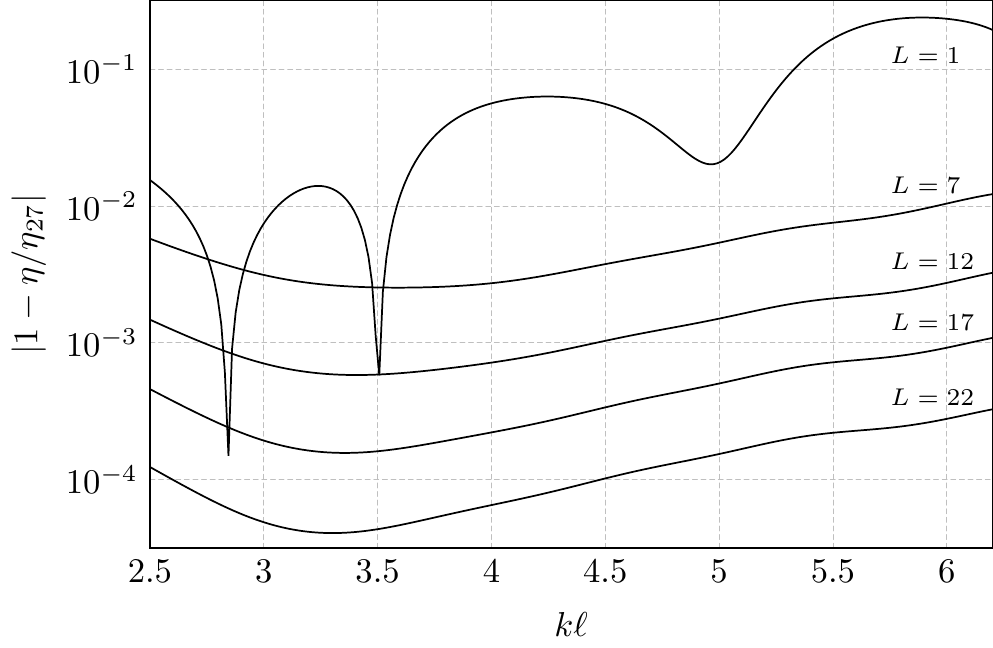}
    \caption{Relative error in radiation efficiency of a dipole in the vicinity of a spherical shell with outer radius~$r=0.8\ell$. Individual lines are computed with a different value of $L$ and compared to the reference~$\eta_{27}$ computed with~$ L = L_\T{max} = 27$ determined from~\eqref{eq:Lmax} for the electric size~$ka\approx31.4$.}
    \label{fig:convergenceL}
\end{figure}

An essential aspect of the calculation using the hybrid method is to correctly determine the necessary number of spherical waves used to reach a given precision. The number of spherical waves is defined as $L$ representing the maximal degree of spherical waves taken into account. In the numerical results above, the number of spherical waves used was~\cite{TayliEtAl_AccurateAndEfficientEvaluationofCMs}
\begin{equation}
    \label{eq:Lmax}
    L = L_\T{max} = \lceil ka+7\sqrt[3]{ka}+3 \rceil
\end{equation}
with~$a$ being the radius of a the smallest sphere centered at origin surrounding both objects. The effect of using smaller values $L < L_\T{max}$ is shown in Fig.~\ref{fig:convergenceL} illustrating the convergence of the radiation efficiency based on the number of spherical waves used.

The above-defined test case can also be adapted to verify the results of the dual definition solution~\eqref{eq:InternalFull}, \ie{}, of the ``internal'' setup. To this point, the dipole antenna is moved to the center of the spherical shell. The size of the dipole is adjusted so that the radius of its smallest circumscribing sphere does not exceed the inner radius of the spherical shell. For the purpose of a comparison of the results with the time-domain solver of the CST Microwave Studio~\cite{cst2021}, the dielectric material forming the spherical body is, in this case, characterized by the real-valued relative permittivity~$\varepsilon_\T{r}$ and electric conductivity~$\sigma$. The compared quantity is reflection coefficient~$\varGamma$ (with reference impedance equal to~$50 \, \Omega$) and radiation efficiency.

\begin{figure}[t]
    \centering
    \includegraphics[width=8.5cm]{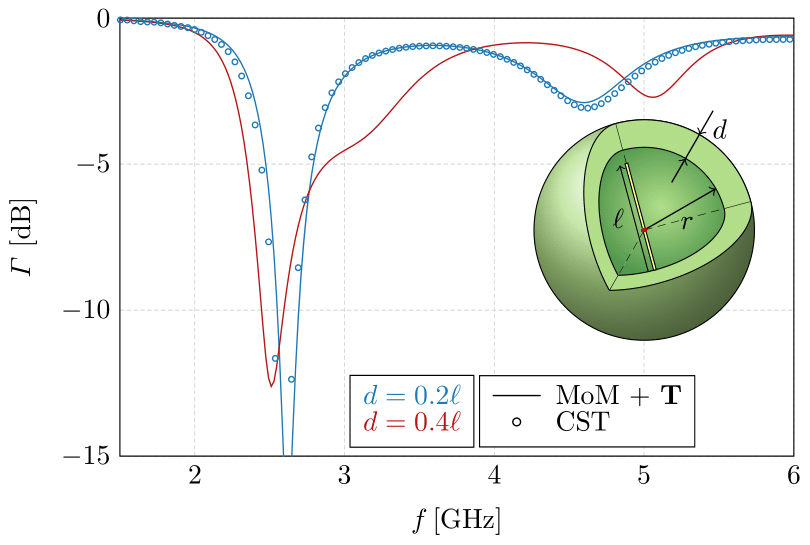}
    \caption{Comparison of reflection coefficient $\varGamma$ seen at the delta-gap feed placed in the middle of the dipole with length $\ell=0.05\,\T{m}$ and an aspect ratio of~$\ell/w=50$, which is encapsulated inside a spherical shell of thickness~$d$ made of a material with relative permittivity~$\varepsilon_\T{r} = 5 $ and conductivity $\sigma=0.1\,\T{S} \cdot \T{m}^{-1}$. The inner radius of the shell is~$r=0.6\ell$. Results for the spherical shell of thickness~$d=0.2\ell$ are displayed in blue and compared with the solution from the time-domain solver of the CST Studio Suite. The red line represents the solution for a shell of thickness~$d=0.4\ell$.}
    \label{fig:T_in_S11}
\end{figure}

Figure~\ref{fig:T_in_S11} shows a comparison of the reflection coefficient~$\varGamma$. As in the previous case, the results of two different numerical schemes are in satisfactory agreement, verifying the method presented in this paper.

\begin{figure}[t]
    \centering
    \includegraphics[width=8.5cm]{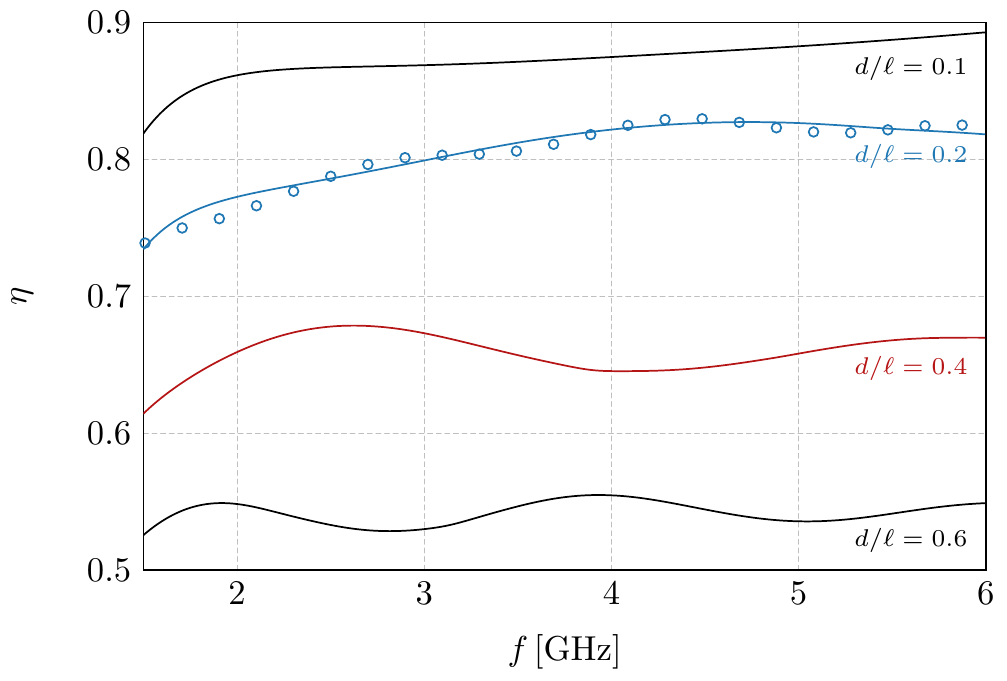}
    \caption{Radiation efficiency $\eta$ of the dipole inside a spherical shell of various thickness $d$ made of the material with relative permittivity~$\varepsilon_\T{r} = 5 $ and conductivity $\sigma=0.1\,\T{S} \cdot \T{m}^{-1}$. Blue and red lines correspond to the same lines from Fig.~\ref{fig:T_in_S11}.}
    \label{fig:T_in_efficiency}
\end{figure}

The second comparison is the same as in the ``external'' case, namely the comparison of radiation efficiency, depending, in this case, on the thickness of~$d/\ell$ of the spherical shell. It can be seen in Fig.~\ref{fig:T_in_efficiency} that the radiation efficiency decreases with the increase of the thickness of the shell. The results were verified against CST for $d/\ell = 0.2$, with good agreement being observed considering the fact that two very different numerical methods were used.

\section{Unification of the External and Internal Formulation}
\label{sec:hybridUnified}

\begin{figure}[t]
    \centering
    \includegraphics[height=4.174cm]{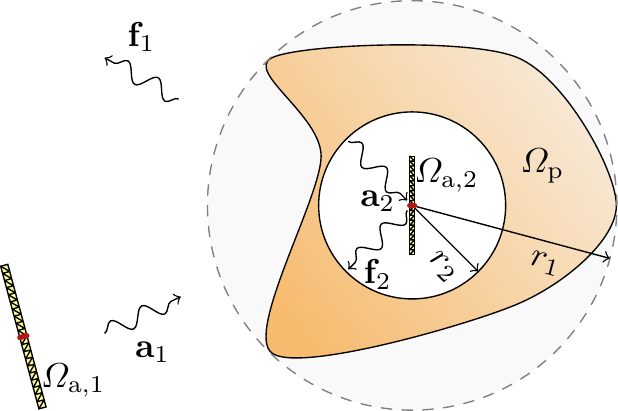}
    \caption{The unified case of the hybrid method that involves a pair of objects described by the \ac{MoM} ($\varOmega_{\T{a},1}$ and $\varOmega_{\T{a},2}$) and one object described with the T-matrix method ($\varOmega_\T{p}$).}
    \label{fig:unified}
\end{figure}

Section~\ref{sec:Hybrid} describes in detail the derivation of the hybrid method for two different cases, the first one with object~$\varOmega_\T{a}$ in the exterior of object~$\varOmega_\T{p}$ and the second with object~$\varOmega_\T{a}$ in the interior cavity carved in object~$\varOmega_\T{p}$. This section combines these two cases. The scenario is depicted in Fig.~\ref{fig:unified} and resembles the communication between two antennas, $\varOmega_{\T{a},1}$ and $\varOmega_{\T{a},2}$, one being in the exterior of a material body, while the second is in its interior, \eg{}, a communication between a reader and an implanted antenna. As in the previous cases, the location of individual objects must fulfill the requirements, specifically that object~$\varOmega_{\T{a},2}$ must fit within a sphere inscribed in object~$\varOmega_\T{p}$ and object~$\varOmega_{\T{a},1}$ cannot enter the circumscribing sphere of object~$\varOmega_\T{p}$.

In order to describe this general scenario, objects~$\varOmega_{\T{a},1}$ and $\varOmega_{\T{a},2}$ are described by impedance matrices~$\M{Z}_1$ and $\M{Z}_2$, respectively, and the spherical wave description of object~$\varOmega_\T{p}$ is generalized to
\begin{equation}
    \label{eq:TmatGeneral}
    \mqty[ \M{f}_1 \\ \M{a}_2] = 
    \mqty[ \M{T} & \M{\Psi} \\
        \M{\Psi}^\T{T} & \M{\Gamma} ]
    \mqty[\M{a}_1 \\ \M{f}_2] ,
\end{equation}
where matrix~$\M{\Psi}$ accounts for the field penetrating from inside of object~$\varOmega_\T{p}$ outwards, while matrix~$\M{\Psi}^\T{T}$ accounts for the field penetrating from outside of object~$\varOmega_\T{p}$ inwards.

The interactions between all three objects $\varOmega_{\T{a},1}$, $\varOmega_{\T{a},2}$, and $\varOmega_\T{p}$ from Fig.~\ref{fig:unified} are described by a system of equations 
\begin{equation}
    \label{eq:unified}
    \mqty[
        \M{Z}_1 & \M{0} & \M{0} & -\utbSmat_4^\T{T} & \M{0} & \M{0} \\
        \M{0} & \M{Z}_2 & -\utbSmat_1^\T{T} & \M{0} & \M{0} & \M{0} \\
        \M{0} & -\utbSmat_1 & \M{0} & \M{0} & \M{0} & -\M{1} \\
        -\utbSmat_4 & \M{0} & \M{0} & \M{0} & -\M{1} & \M{0} \\
        \M{0} & \M{0} & \M{0} & -\M{1} & \M{T} & \M{\Psi} \\
        \M{0} & \M{0} & -\M{1} & \M{0} & \M{\Psi}^\T{T} & \M{\Gamma}]
    \mqty[\M{I}_1 \\ \M{I}_2 \\ \M{a}_2 \\ \M{f}_1 \\ \M{a}_1 \\ \M{f}_2] = \mqty[\M{\widetilde{V}}_1^\T{i}\\ \M{V}_2^\T{i} \\ \M{0} \\ -\M{a}^\T{i} \\ \M{0} \\ \M{0}],
\end{equation}
where $\M{\widetilde{V}}_1^\T{i} = \M{V}_1^\T{i} + \utbSmat_1^\T{T}\M{a}^\T{i}$. The definitions \eqref{eq:ExternalFull} and~\eqref{eq:InternalFull} are then only the special cases of this unified definition.
When only the interaction of currents~$\M{I}_1$ and $\M{I}_2$ is required, the system~\eqref{eq:unified} can partially be resolved, leading to the equation system
\begin{equation}
    \label{eq:unifiedI1I2}
\mqty[\M{Z}_1+\utbSmat_4^\T{T}\M{T}\utbSmat_4 & \utbSmat_4^\T{T}\M{\Psi}\utbSmat_1  \\
            \utbSmat_1^\T{T}\M{\Psi}^\T{T}\utbSmat_4 & \M{Z}_2 + \utbSmat_1^\T{T}\M{\Gamma}\utbSmat_1]
\mqty[\M{I}_1 \\ \M{I}_2] = 
\mqty[\M{\widetilde{V}}_1^\T{i} + \utbSmat_4^\T{T} \M{T} \M{a}^\T{i}\\ \M{V}_2^\T{i} + \utbSmat_1^\T{T}\M{\Psi}^\T{T}\M{a}^\T{i}].
\end{equation}

\section{Results: Application}
\label{Sec:ResultsII}
The previous section revealed how the external~\eqref{eq:ExternalFull} and internal~\eqref{eq:InternalFull} formulation can be interlinked forming a general formulation~\eqref{eq:unified}. This section builds on this definition and shows its use in a simplified electromagnetic problem representing communication between two antennas, one of them being implanted in the human head and the second in its exterior. Models of this kind have previously been employed in the evaluation of fundamental bounds on implantable antennas~\cite{Skrivervik_Bosiljevac_Sipus_FundamentalBoundsForImplantedAntennas} which is a field of research where the presented hybrid method can offer substantial advantages.

To highlight the benefits of the presented method, the human head is modeled as a spherical multi-layer, the parameters of which are specified in Table~\ref{tab:head}. Material properties of different human tissues are obtained from~\cite{DielectricPropertiesofBodyTissues}.
\begin{table}[t]
    \label{tab:head}
    \centering
    \caption{Structure of the layer model of the human head~\cite{Drossos_TheDependeceofElectromagneticEnergyAbsorptionUponHumanHeadTissues}.}
    \begin{tabular}{lrr}
        Layer & Thickness & Density \\ \toprule
        Brain white matter & $71.0\,\T{mm}$ & $1041\,\T{kg}\cdot\T{m}^{-3}$\\ 
        Brain grey matter & $10.0\,\T{mm}$ & $1045\,\T{kg}\cdot\T{m}^{-3}$\\
        Bone & $6.6\,\T{mm}$ & $1908\,\T{kg}\cdot\T{m}^{-3}$\\
        Fat & $1.4\,\T{mm}$ & $911\,\T{kg}\cdot\T{m}^{-3}$\\
        Skin & $1\,\T{mm}$ & $1109\,\T{kg}\cdot\T{m}^{-3}$\\ \bottomrule
    \end{tabular}
\end{table}

The spherical multi-layer is centered at the origin and has a  spherical cavity of radius~$14.5\,\T{mm}$ in which a \ac{PEC} strip dipole antenna with length $\ell=25\,\T{mm}$ and aspect ratio of~$\ell/w=50$ is placed. Another \ac{PEC} strip dipole antenna of the same dimensions is placed in the exterior of the multi-layer model at distance~$20\,\T{mm}$ from the outer surface of the head (the normal distance to the surface of the dipole). Both dipole antennas are fed in their geometric center by a delta-gap feed and their spatial orientation is the same. Both antennas are further supplemented with L-shaped impedance matching circuits~\cite{Pozar_MicrowaveEngineering} consisting of ideal lumped capacitors and inductors. Two sets of results presented in this section differ only in the matching frequency, which is~$5\,\T{GHz}$ or~$7\,\T{GHz}$. Matching is realized with respect to terminal impedance~$50\,\Omega$.

Communication of two antennas in a complex environment can be represented as a two-port network~\cite{Pozar_MicrowaveEngineering}, in which delta-gap feeds of both antennas, together with matching circuits, form the ports. Scattering parameters~\cite{Pozar_MicrowaveEngineering} are used to describe the network.

\begin{figure}[t]
    \centering
    \includegraphics[width=8.5cm]{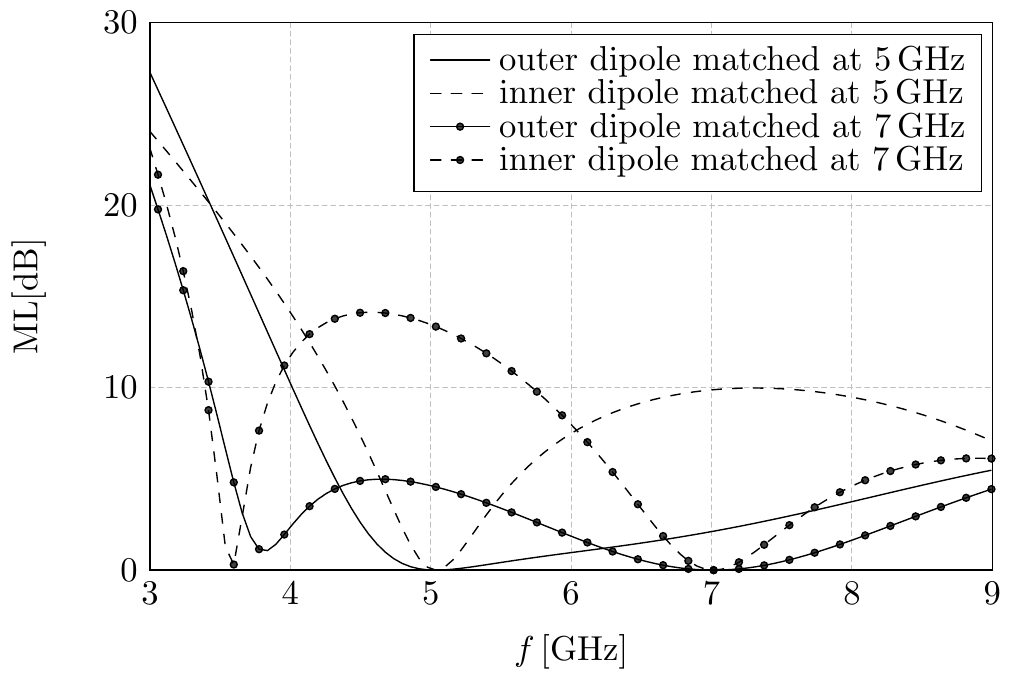}
    \caption{The mismatch loss~$\T{ML}$ at the ports of the two-dipole system. Bare lines correspond to the matching at a frequency equal to~$5\,\T{GHz}$. Lines with markers belong to the second case with impedance matching at a frequency equal to~$7\,\T{GHz}$. Matching is performed with respect to an impedance of $50\, \Omega$. }
    \label{fig:head_mismatchLoss}
\end{figure}

\begin{figure}[t]
    \centering
    \includegraphics[width=8.5cm]{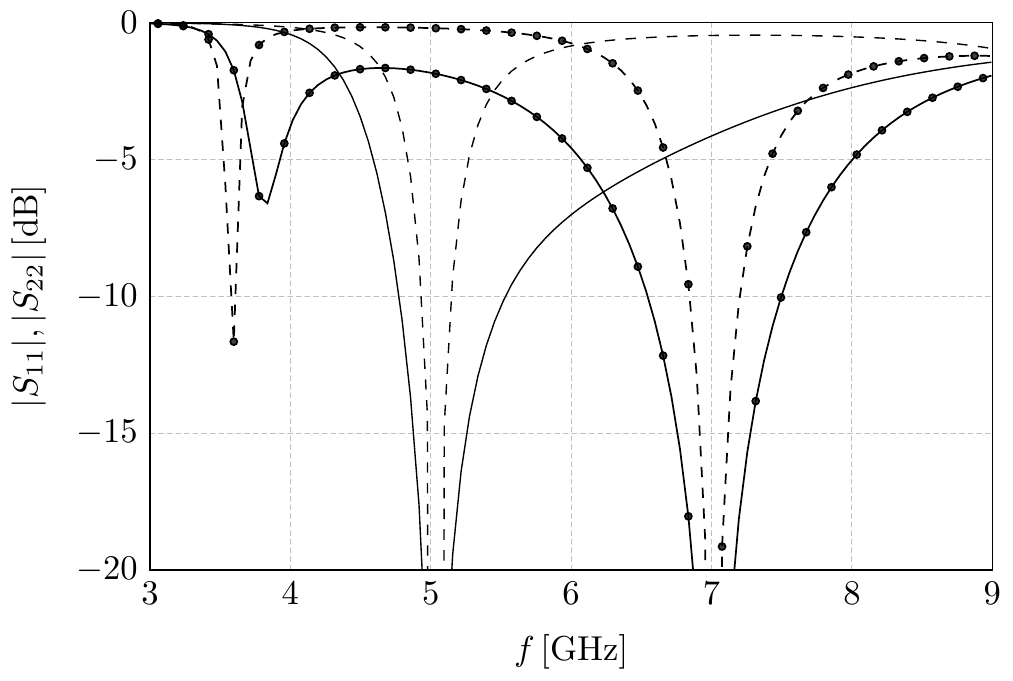}
    \caption{Reflectances seen at ports of the system for the same setup as in~Fig.~\ref{fig:head_mismatchLoss}.}
    \label{fig:head_refflection}
\end{figure}

Mismatch loss \mbox{$\T{ML} = - 10 \T{log}_{10} \left( 1 - \lvert S_{ii} \rvert^2 \right)$} and reflectances~$\lvert S_{ii} \rvert^2$ are shown in Fig.~\ref{fig:head_mismatchLoss} and Fig.~\ref{fig:head_refflection}, respectively, assuming ports with a characteristic impedance of $50\, \Omega $.

\begin{figure}[t]
    \centering
    \includegraphics[width=8.5cm]{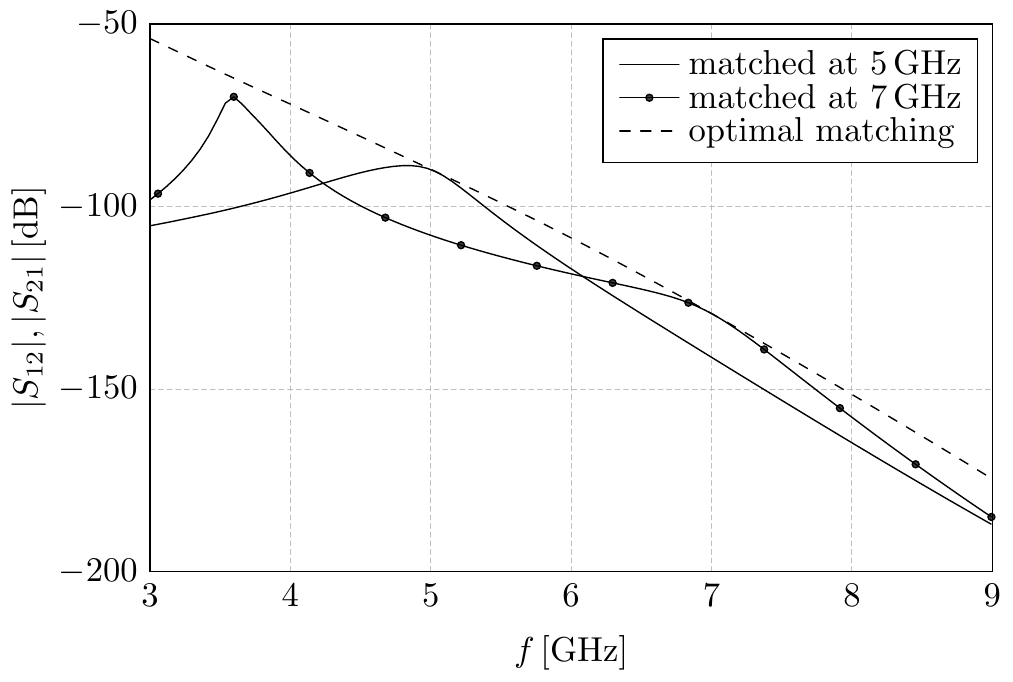}
    \caption{Trasmittances $\lvert S_{12}\rvert^2$ and $\lvert S_{21}\rvert^2$ as a function of frequency in the system used in~Fig.~\ref{fig:head_refflection}. Full lines represent matching at a frequency of $5\,\T{GHz}$ or $7\,\T{GHz}$. The dashed line corresponds to transmittance when perfect matching is provided at every frequency.}
    \label{fig:head_mutualS}
\end{figure}

The effect of matching on the transmittance of the system is shown in Fig.~\ref{fig:head_mutualS}. It can be observed that the frequency dependence of transmittance differs qualitatively before and after the matching frequency. At frequencies below the matching frequency, the impedance mismatch, seen in Fig.~\ref{fig:head_mismatchLoss}, highly affects the transmittance between the antennas. On the contrary, at higher frequencies, the effect of the impedance mismatch is weaker. In this frequency region, the transmittance is mostly dictated by the lossy multi-layer (human head model). This claim is supported by the dashed curve in Fig.~\ref{fig:head_mutualS}, which shows transmittance for the case of perfect matching at every frequency.

\begin{figure}[t]
    \centering
    \includegraphics[height=6.57cm]{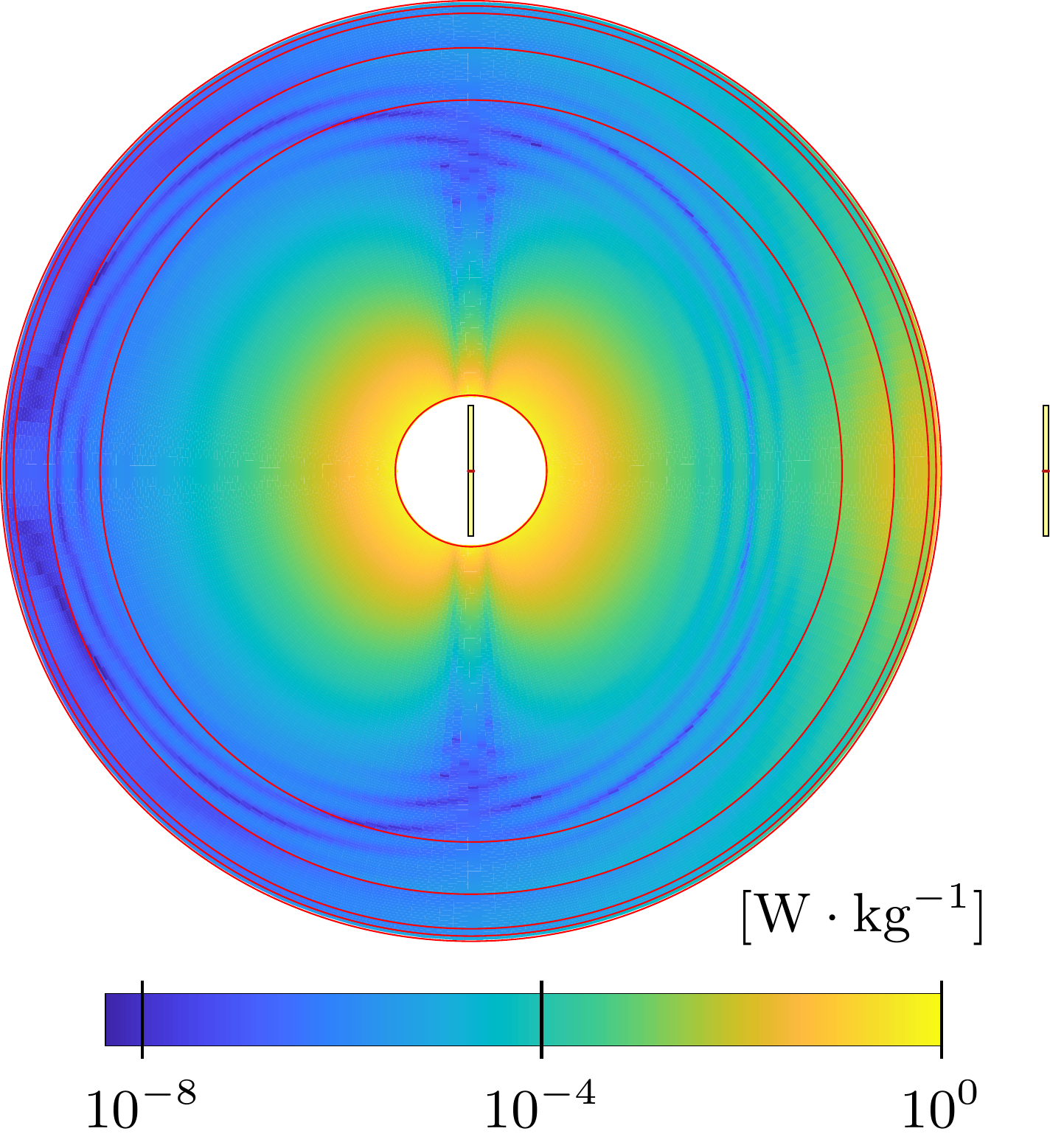}
    \caption{Specific absorption rate at frequency $f = 5\, \T{GHz}$ computed for the multi-layer model of the human head. The cycle mean input power to the system is set to~$10\,\T{mW}$, where~$8.1\,\T{mW}$ belongs to the internal dipole and the rest to the external dipole.}
    \label{fig:head_SAR}
\end{figure}

When the human body is exposed to a radio frequency electromagnetic field, the specific absorption rate (SAR)
\begin{equation}
    \T{SAR} = \dfrac{\sigma \left| \V{E} \right|^2}{2 \rho},
\end{equation}
where~$\sigma$ is conductivity and~$\rho$ is the mass density, becomes another metric of interest~\cite{GuidelinesforLimitingExposuretoElectromagneticFields}. In the spherical multi-layer scenario described in this section, the evaluation of this quantity is a straightforward task the result of which is shown in Fig.~\ref{fig:head_SAR} for the case when both dipoles are excited by the same voltage across the delta-gap feed. The cycle mean input power to the system has been set to~1\,W.

\begin{figure}[t]
    \centering
    \includegraphics[width=8.5cm]{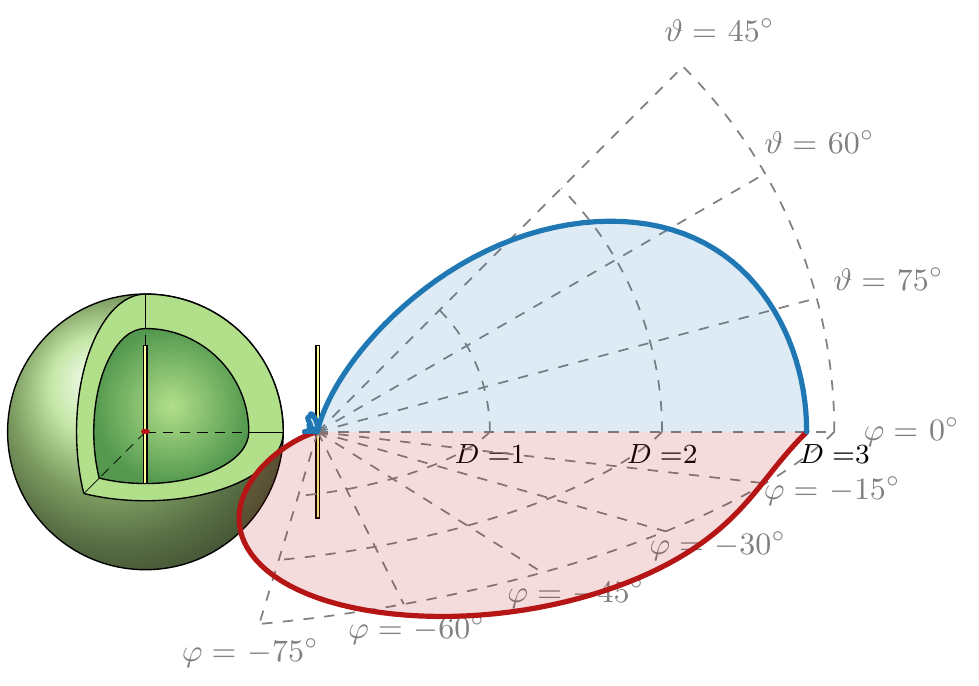}
    \caption{Two principal cuts of a directivity radiation pattern at $5\,\T{GHz}$ for the spherical multi-layer setup excited solely by the external dipole antenna. Partial directivity with polarization along~$\vartheta$~direction is shown.}
    \label{fig:head_farfield}
\end{figure}

Another quantity of interest is the radiation pattern. The most important one being the radiation pattern of the outer radiator which will be greatly affected by the presence of the conducting spherical multi-layer. In order to evaluate the electric far field
\begin{equation}
    \label{eq:farzone}
    \V{F}\left(\vartheta, \varphi\right)=\lim_{r\rightarrow\infty}\{r\T{e}^{\T{j}kr}\V{E}\left(\V{r}\right)\},
\end{equation}
it is important to realize that the total the electric field is produced by waves~$\M{f}_1$ emanating from object~$\varOmega_\T{p}$ and field~\eqref{eq:Es} produced by object~$\varOmega_\T{a}$. Since spherical wave decomposition is already employed in the formulation, it can also be advantageously used to express the electric far field. Notably, the field~\eqref{eq:Es} is transformed into out-going spherical waves~$\M{f}_{\V{J}} = - \utbSmat_1 \M{I}$ and is afterwards summed with waves~$\M{f}_1$. The resulting electric far field is written as
\begin{equation}
    \V{F}\left(\vartheta, \varphi\right) = \sqrt{Z}\sum_\alpha\left(f_{\V{J}\alpha} + f_{1\alpha}\right)\T{j}^{l+2-\tau}\M{Y}_\alpha\left(\widehat{\V{r}}\right),
\end{equation}
where~$\M{Y}_\alpha$ are the vector spherical harmonics described in~Appendix~\ref{App:sphericalExpansion}. The radiation pattern generated by the studied setup, when only the external dipole is excited, is presented in Fig.~\ref{fig:head_farfield}.

\section{Results: Computational Efficiency}

Computational efficiency is an important aspect of every numerical method and is detailed in this section for the proposed hybrid. Two scenarios are addressed. The first scenario consists of a dipole inside a spherical shell that was already used in Section~\ref{Sec:ResultsI:HybridMethod}. The second is a demanding setup of two dipoles communicating through a dielectric cube.

The computational times for the case of a dipole inside a spherical shell are shown in Table~\ref{tab:tabSphShellDipole}. The table shows a comparison of total computation time with CST time-domain solver and also presents evaluation times for different parts of the hybrid scheme. Concerning the hybrid scheme, it can be stated that the time needed for the build-up and solution of the equation system~\eqref{eq:InternalI} is negligible as compared to the previous stages. In this particular and favorable scenario, it can also be stated that the construction of matrix~$\M{\Gamma}$ is a minor burden even for the very high number of spherical waves~$L_\T{max} = 17$ since the evaluation of matrix~$\M{\Gamma}$ is analytical for a spherical multi-layer. The computation time is therefore dominated by the construction of matrix~$\M{Z}$ of the internal dipole and by the construction of coupling matrices~$\M{U}$, the evaluation time of which also scales with the number of spherical waves. The solution of the total system is equivalent to a few solutions to the dipole radiating in free space. When frequency sweep in the range from~$1.5 \, \T{GHz}$ to $6 \, \T{GHz}$ is demanded, the evaluation time using CST solver is equivalent to approximately $500$~runs of the hybrid method for~$L_\T{max} = 11$ (corresponding to the lowest frequency) and $350$~runs for~$L_\T{max} = 17$ (corresponding to the highest frequency). This can be considered as great computational efficiency realizing that the computational time is almost independent of the material composition of the spherical multi-layer.

\begin{table}[!h]
    \caption{Computational times for a dipole inside a spherical shell}
    \centering
    \begin{tabular}{lcc}
        evaluated task & \multicolumn{2}{c}{computational time} \\ \toprule
        CST time-domain solver & \multicolumn{2}{c}{$1700$\,s} \\ \toprule
        hybrid & $L_\T{max} = 11$ & $L_\T{max} = 17$ \\ 
         & $\left(ka = 0.79\right)$ & $\left(ka = 3.14\right)$ \\\midrule
        \quad \textbullet\, dipole $\M{Z}$ & \multicolumn{2}{c}{$1.8\,\mathrm{s}$} \\
        \quad \textbullet\, dipole $\mathbf{U}_1$ & $1.2\,\mathrm{s}$ & $2.1\,\mathrm{s}$ \\
        \quad \textbullet\, spherical shell $\mathbf{\Gamma}$ & $0.32\,\mathrm{s}$ & $0.80\,\mathrm{s}$ \\
        \quad \textbullet\, hybrid solution & $0.024\,\mathrm{s}$ & $0.035\,\mathrm{s}$ \\ \midrule
        hybrid total time & $3.4\,\mathrm{s}$ & $4.8\,\mathrm{s}$ \\ \bottomrule
    \end{tabular}
    \label{tab:tabSphShellDipole}
\end{table}

\begin{figure}[t]
    \centering
    \includegraphics[width=8.5cm]{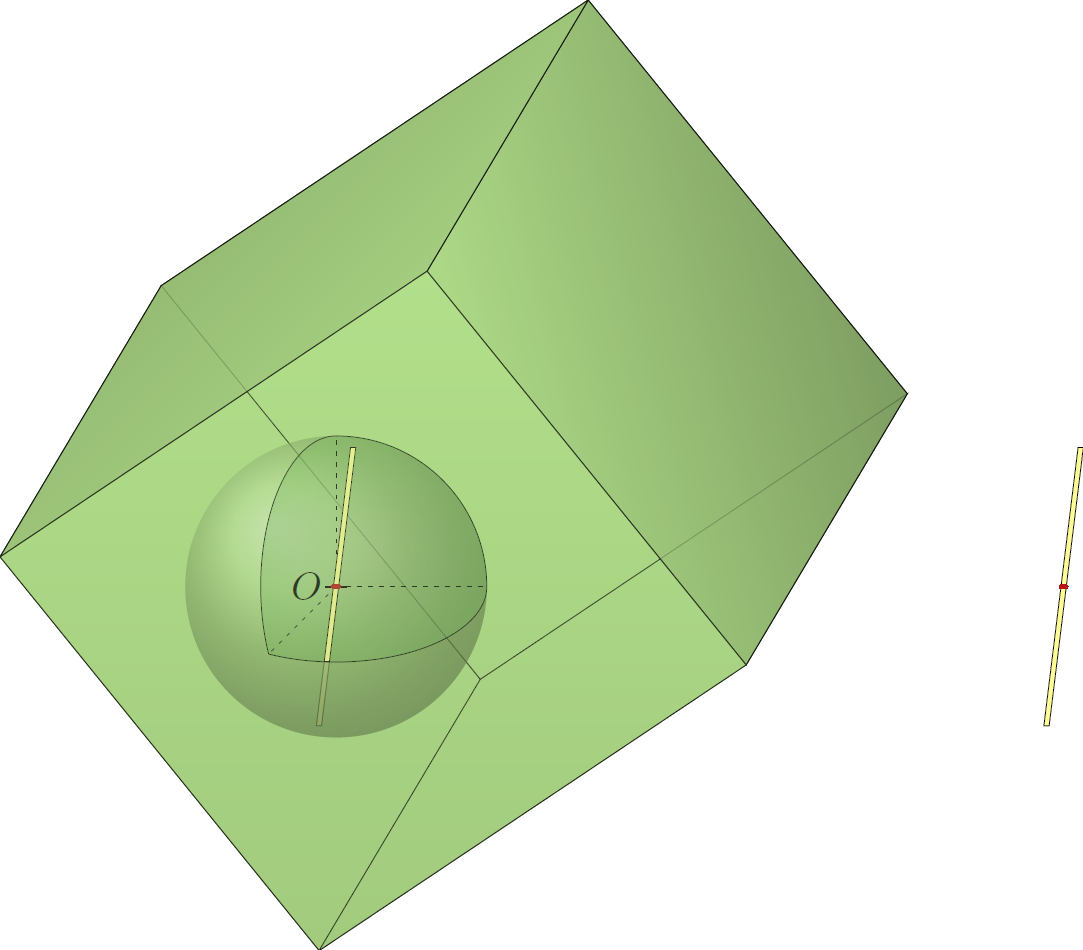}
    \caption{Spatial arrangement of the two dipoles, one of which is placed in a vacuum bubble inside a dielectric cube.}
    \label{fig:twoDipolesCube}
\end{figure}

The scenario of two dipoles and a cube is sketched in Fig.~\ref{fig:twoDipolesCube} and represents a computationally demanding setup for the hybrid method. This non-trivial scenario is used to analyze the convergence and computational complexity of the calculations depending on the value of $L$. 

\begin{figure}[t]
    \centering
    \includegraphics[width=8.5cm]{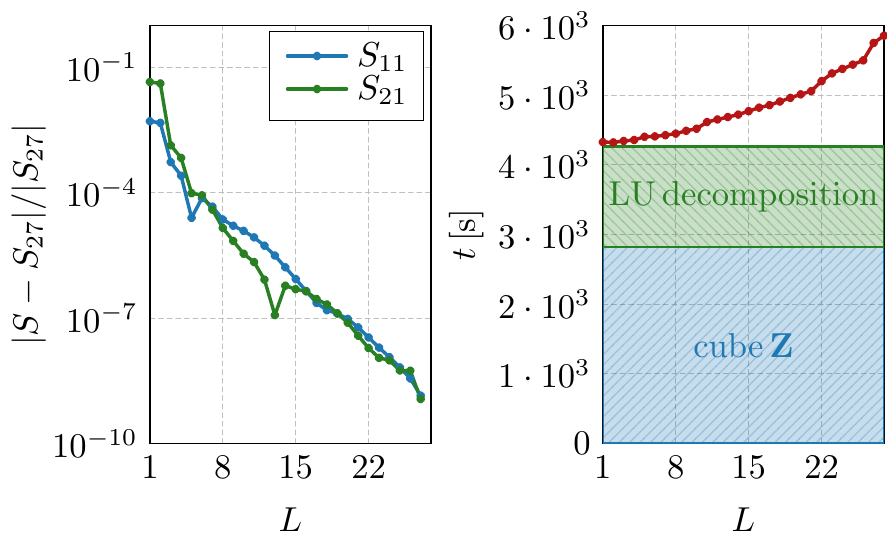}
    \caption{Relative error in scattering parameters as seen from the two delta-gap ports placed at the center of each dipole from Fig.~\ref{fig:twoDipolesCube} and the corresponding computational times. The electrical size of the cube is $ka \approx 6.7$. The individual points are calculated with different values of $L$ and their relative deviation is then determined by comparison with $S_{27}$ calculated with $L_\mathrm{max} = 27$, which was determined using \eqref{eq:Lmax}. The main component of the calculation time is the determination of the impedance matrix and its LU decomposition. The scaling with the number of spherical waves is not significant.}
    \label{fig:cubeLconvergenceTime}
\end{figure}

As can be seen from Table~\ref{tab:twoDipolesCube} and Fig.~\ref{fig:cubeLconvergenceTime}, the major computational burden here is the construction of matrices~$\M{T}$, $\M{\Gamma}$, and $\M{\Psi}$ which requires the construction of the impedance matrix of the cube and its LU decomposition for the evaluation of~\eqref{eq:TmatToZ}. For homogeneous obstacles, this burden can be reduced by the use of formulation based on surface equivalence~\cite{Jin_TheoryAndComputationOfElectromagneticFields} but will still be considerable realizing that the electrical size of the cube is~$ka \approx 1$ at the lowest frequency and~$ka \approx 12$ at the highest frequency and that the cube is made of a dielectric with relative permittivity~$\varepsilon_\T{r} = 2$.  Figure.~\ref{fig:cubeLconvergenceTime} also shows that even using only the dominant modes ($L = 1$) it is possible to obtain relative error of~$10^{-1}$ and that the use of a larger number of spherical waves does not lead to a significant increase in computational cost. 

For a single evaluation of this challenging scenario, the proposed hybrid method cannot compete with computational schemes offered by CST time-domain solved or by FEM+MoM hybrid offered by FEKO. The hybrid method can nevertheless be advantageous in scenarios when different positions/orientations of the dipoles are to be studied. In such case, only the coupling matrices~$\M{U}$ and the final hybrid solution must be recalculated for every new position/orientation of the dipoles unlike in the case of the used commercial solvers.

\begin{table}[!h]
    \centering
    \caption{Computational times for the setup sketched in Fig.~\ref{fig:twoDipolesCube}}
    \begin{tabular}{lcc}
        evaluated task & \multicolumn{2}{c}{computation times} \\ \toprule
        CST time-domain solver & $310$\,s & total time \\ 
        FEKO FEM+MoM & $470$\,s & one frequency point\\ \toprule
        hybrid & $L_\T{max} = 14$ & $L_\T{max} = 39$ \\ 
        & $\left(ka = 1.72\right)$ & $\left(ka = 17.29\right)$ \\ \midrule
        
        \quad\textbullet\, dipole MoM & $3.2\,\mathrm{s}$ & $3.2\,\mathrm{s}$ \\
        \quad\textbullet\, dipole ext. $\mathbf{U}_1$/$\mathbf{U}_4$ & $1.9\,\mathrm{s}/2.6\,\mathrm{s}$ & $12\,\mathrm{s}/16\,\mathrm{s}$ \\
        \quad\textbullet\, dipole int. $\mathbf{U}_1$/$\mathbf{U}_4$  & $1.6\,\mathrm{s} / 2.7\,\mathrm{s}$ & $9.7\,\mathrm{s} / 14\,\mathrm{s}$ \\
        \quad\textbullet\, cube MoM3D & $2500\,\mathrm{s}$ & $2500\,\mathrm{s}$ \\
        \quad\textbullet\, cube $\mathbf{U}_1$/$\mathbf{U}_4$ & $78.0\,\mathrm{s}/180\,\mathrm{s}$ & $580\,\mathrm{s}/1000\,\mathrm{s}$ \\
        \quad\textbullet\, cube $\mathbf{T}$ & $1400\,\mathrm{s}$ & $1524\,\mathrm{s}$ \\
        \quad\textbullet\, cube $\mathbf{\Gamma}$ + $\mathbf{\Psi}$ & $1400\,\mathrm{s}$ & $1500\,\mathrm{s}$ \\
        \quad\textbullet\, hybrid solution & $0.15\,\mathrm{s}$ & $0.54\,\mathrm{s}$ \\ \midrule
        hybrid total time & $5500\,\mathrm{s}$ & $7200\,\mathrm{s}$ \\ \bottomrule
    \end{tabular}
    \label{tab:twoDipolesCube}
\end{table}

\section{Discussion}
\label{sec:Discussion}
Sections~\ref{Sec:ResultsI} and~\ref{Sec:ResultsII} illustrate the numerical performance of the method and the agreement of the results with solutions given by state-of-the-art commercial simulators. This section intends to discuss the strengths and weaknesses of the technique.

The major strength of the proposed computational scheme is the separated treatment of a complex, but electrically small object, represented by the \ac{MoM} matrix~$\M{Z}$ and of a typically simple, but electrically large body, described by its matrix~$\M{T}$. An essential feature connected with this separation is independence, in terms of how matrix~$\M{T}$ is determined. This, in many cases, allows the scatterer~$\varOmega_\T{p}$ to be described using an analytic formula, which is, for example, the case of generic spherical, cylindrical, and planar multi-layer obstacles~\cite{Stratton_ElectromagneticTheory, Kristensson_ScatteringBook} speeding up the entire hybrid method. The second essential property emanating from the separation of two interacting objects is the favorable scaling with the electrical size of object~$\varOmega_\T{p}$. The computational complexity depends on the number of spherical waves used which is equal to~$2L \left(L + 2 \right)$ which, for large electric sizes, is proportional to~$\left(ka \right)^2$. In contrast, classical computational schemes using the segmentation of the body into mesh cells typically operate with $\left(ka\right)^3$ scaling. In that respect, the internal scenario described in Sec.~\ref{sec:HybridInt} with matrix~$\M{T}$ given analytically is the most favorable as there the necessary number of spherical waves is dictated by object~$\varOmega_\T{a}$. The computational burden is then independent of the electrical size of object~$\varOmega_\T{p}$. Finally, the method excels in scenarios when object~$\varOmega_\T{p}$ is fixed and the various positioning of object~$\varOmega_\T{a}$ are investigated. In such a case, the computationally demanding parts, matrices~$\M{Z}$ and~$\M{T}$, are calculated once and only the coupling matrix~$\utbSmat_1$ is iteratively recalculated.

One of the important outcomes of this paper are relations \eqref{eq:ExternalI}, \eqref{eq:InternalI}, and~\eqref{eq:unifiedI1I2} containing the reduced description of the treated setup taking the point of view of object~$\varOmega_\T{a}$ and resembling the use of Green's function in the presence of boundary conditions induced by object~$\varOmega_\T{p}$~\cite{Dudley_MathematicalFoundationsForElectromagneticTheory}. This leads to a reduction of memory and inversion demands but still leaves the possibility of operating with the complete system matrix allowing for the formulation of fundamental bounds on antenna and scattering metrics in the presence of parasitic scatterer residing in region~$\varOmega_\T{p}$. This reduced description also removes the unpleasant numerical dynamics of transition matrix components corresponding to high-order spherical waves via the favorable multiplication by matrix~$\utbSmat_4$. This last step can, in many cases, be pre-calculated, straightening even more the numerical robustness.

An important side product of the development of the hybrid method is an explicit formula to evaluate matrix~$\M{T}$ from matrix~$\M{Z}$ via relations~\eqref{eq:TmatToZ}, which generalizes previously published results~\cite[(12)]{2013_Kim_TAP} and~\cite[(20)]{2017_Markkanen_JQSRT}. Though this method is computationally intensive, it is numerically stable, even in cases when null-field exhibits instabilities~\cite{Mishchenko_ScatteringAbsorptionandEmissionogLightbySmallParticles}. The versatility of this approach was documented on an example of the \ac{PEC} cube. it might be the only possibility for complex shaped objects. An important outcome of this connection is also a possibility to evaluate characteristic numbers of arbitrarily shaped objects with fourfold numerical dynamics as opposed to the classical approach~\cite{HarringtonMautz_ComputationOfCharacteristicModesForConductingBodies,HarringtonMautzChang_CharacteristicModesForDielectricAndMagneticBodies}. Considering these advantages and readily available codes based on the method of moments formulation of field integral equations, relations~\eqref{eq:TmatToZ} should become a standard way to evaluate matrices~$\M{T}$, $\M{\Psi}$, and~$\M{\Gamma}$.

As is the case of every numerical technique, the proposed method also exhibits weaknesses. An important one is the assumption that matrix~$\M{T}$ of an electrically large object can be obtained in a numerically effective way. This might be problematic for complex material objects since it requires solution to a linear equation system, see~\eqref{eq:TmatToZ}, in cases when analytical solution is not available. Another weakness is the assumption that the number of necessary spherical waves to achieve satisfactory precision is not exceedingly high. This forbids scenarios when near field interaction between the object~$\varOmega_\T{p}$ and object~$\varOmega_\T{a}$ dominates, as is the case of an antenna put in contact with a human head, or scenarios in which the object~$\varOmega_\T{p}$ is considerably off-centered as compared to the wavelength. Another limitation is the assumption that object~$\varOmega_\T{a}$ must lie entirely outside a circumscribed sphere or inside an inscribed sphere of object~$\varOmega_\T{p}$. This prevents us from studying situations in which object~$\varOmega_\T{a}$ is positioned in protrusions of object~$\varOmega_\T{p}$ as in, for example, the case of antennas mounted on electrically large carriers. 

\section{Conclusion}
\label{Sec:Conclusion}

A hybrid of the \ac{MoM} and the T-matrix method was introduced combining the piecewise-defined basis functions approximating electrically small, however, potentially highly irregular objects, and spherical waves representing the scattering properties of a potentially electrically large obstacle of a regular shape. The presence of an inhomogeneous material is taken into account. The resulting system matrix, of numerically tractable size, is in most cases, suitable for finding a direct solution. As long as the electrically large object remains unchanged, the method allows for the efficient recalculation of only a low-rank coupling matrix when the positioning of small objects is of interest. 

The method is based on a set of projection matrices which interconnect piecewise-defined and entire-domain representations. The side-product of the method is a simple formula for the determination of the transition matrix from the impedance matrix describing an arbitrary object. The projection matrices used can be utilized for characteristic mode decomposition quadruplicating the numerical dynamics of the conventional methods.

The hybrid method has been constructed for scenarios such as communication of an implanted antenna with an external reader including the exposure evaluation. Since the system matrix is explicitly determined and stored in a computer's memory, the method makes it possible to employ eigenvalue methods and determine fundamental bounds on a variety of physical phenomena. This is a subject of ongoing research.

\appendices
\section{Spherical Vector Waves Expansions}
\label{App:sphericalExpansion}
The spherical vector waves are defined as triplets ($\tau = 1, 2, 3$) of functions $\M{u}_{\tau\sigma ml}^{(p)}\left(k\V{r}\right) = \M{u}_{\tau\beta}^{(p)}\left(k\V{r}\right)$ \cite{Kristensson_ScatteringBook}, where only two of these are needed for the purpose of this paper. Namely, 
\begin{equation}
    \begin{array}{rcl}
        \M{u}_{1\beta}^{(p)}\left(k\V{r}\right) & = & \T{R}_{1l}^{\left(p\right)}\left(kr\right) \M{Y}_{1\beta}\left(\widehat{\V{r}}\right), \\
        \M{u}_{2\beta}^{(p)}\left(k\V{r}\right) & = & \T{R}_{2l}^{\left(p\right)}\left(kr\right)\M{Y}_{2\beta}\left(\widehat{\V{r}}\right) +  \T{R}_{3l}^{\left(p\right)}\left(kr\right)\M{Y}_{3\beta}\left(\widehat{\V{r}}\right),
    \end{array}
\end{equation}
where $\V{r}$ is the radius vector with relations $\widehat{\V{r}} = \V{r}/{r}$ and~$r = \left| \V{r} \right|$. Multi-index $\beta$ combines indices $l\in \{1,\dots,L\}$ (degree), $m\in\{0,\dots,l\}$ (order) and $\sigma=\{\T{even},\T{odd}\}$ (parity). Function $\T{R}_{\tau l}^{\left(p\right)}$ specifies the radial function~\cite{Hansen_SphericalNearFieldAntennaMeasurements} as
\begin{equation}
\label{eq:AppA:Rfunc}
    \begin{array}{rcl}
        \T{R}_{1l}^{\left(p\right)}\left(kr\right) & = & \T{z}^{\left(p\right)}_l\left(kr\right), \\
        \T{R}_{2l}^{\left(p\right)}\left(kr\right) & = & \dfrac{\left(kr\,\T{z}^{\left(p\right)}_l\left(kr\right)\right)'}{kr}, \\
        \T{R}_{3l}^{\left(p\right)}\left(kr\right) & = & \sqrt{l\left(l+1\right)}\,\dfrac{\T{z}^{\left(p\right)}_l\left(kr\right)}{kr},
    \end{array}
\end{equation}
in which $\T{z}_l^{\left(p\right)}$ represents a spherical Bessel function of degree $l$ and the choice of its variation by superscript $p$ determines the type of the wave. Regular waves $\T{z}^{\left(1\right)}_l = \T{j}_l$ are given by the spherical Bessel function of the first kind, and out-going waves $\T{z}_l^{\left(4\right)} = \T{h}^{\left(2\right)}_l$ are given by the spherical Hankel function of the second kind. The vector spherical harmonics $\M{Y}_\beta$ are defined as
\begin{equation}
    \begin{array}{rcl}
         \M{Y}_{1\beta}\left(\widehat{\V{r}}\right) & = & \dfrac{1}{\sqrt{l\left(l+1\right)}}\nabla\times\left(\V{r}\T{Y}_\beta\left(\widehat{\V{r}}\right)\right), \\
         \M{Y}_{2\beta}\left(\widehat{\V{r}}\right) & = & \widehat{\V{r}}\times\M{Y}_{1\beta}\left(\widehat{\V{r}}\right), \\
         \M{Y}_{3\beta}\left(\widehat{\V{r}}\right) & = & \widehat{\V{r}}\T{Y}_\beta\left(\widehat{\V{r}}\right),\\
    \end{array}
\end{equation}
in which $\T{Y}_\beta$ stands for the spherical harmonic
\begin{equation}
    \T{Y}_{\sigma lm} = \sqrt{\dfrac{2 - \delta_{m0}}{2\pi}}\widetilde{\T{P}}_l^m\left(\T{cos}\vartheta\right)\bigg\{
    \begin{array}{c}
        \T{cos}\left(m\varphi\right)\\
        \T{sin}\left(m\varphi\right)
    \end{array}
    \bigg\}
\end{equation}
with $\widetilde{\T{P}}_l^m$ being the normalized associated Legendre polynomial of degree $l$ and $\delta_{mn}$ being the Kronecker delta.

The spherical vector waves can be used to expand the dyadic Green's function for an electric field \cite{Hansen_SphericalNearFieldAntennaMeasurements, Kristensson_ScatteringBook} as
\begin{equation}
    \label{eq:GreenExp}
    \V{G}_\T{e}\left(\V{r}_1,\V{r}_2\right)=-\T{j}k\sum_{\alpha}\M{u}_{\alpha}^{\left(1\right)}\left(k\V{r}_<\right)\M{u}_{\alpha}^{\left(4\right)}\left(k\V{r}_>\right),
\end{equation}
where $\V{r}_{</>} = \V{r}_{1/2}$ if $r_1 < r_2$ and $\V{r}_{</>} = \V{r}_{2/1}$ if $r_1 > r_2$ and multi-index $\alpha$ combines indices $\tau$ and $\beta$. A possible ordering for the multi-index $\alpha$ can be found in\cite[(7)]{TayliEtAl_AccurateAndEfficientEvaluationofCMs}. Substituting expansion~\eqref{eq:GreenExp} into~\eqref{eq:Es} leads to
\begin{equation}
   \V{E}^\T{s}\left(\V{r}_1\right) = - k^2 Z \left\langle \sum_{\alpha}\M{u}_{\alpha}^{\left(1\right)}\left(k\V{r}_<\right)\M{u}_{\alpha}^{\left(4\right)}\left(k\V{r}_>\right), \V{J} \left(\V{r}_2\right) \right \rangle.
\end{equation}
Assuming that observation point~$\V{r}_1$ is outside of a sphere circumscribing source~$\V{J}$ and substituting~\eqref{eq:Jexp}, the relation is simplified to
\begin{equation}
\label{eq:EsExp}
   \V{E}^\T{s}\left(\V{r}_1\right) = - k \sqrt{Z}  \sum_{\alpha} \left[ \utbSmat_{1} \M{I}\right]_\alpha \M{u}_{\alpha}^{\left(4\right)}\left(k\V{r}_1\right),
\end{equation}
where projector~$\utbSmat_p$ is defined in~Section~\ref{sec:Hybrid}.
If, on contrary, there is a spherical cavity centered at origin and inscribed to source~$\V{J}$, and if the observation point is in the interior of this cavity, the analogous relation reads
\begin{equation}
   \V{E}^\T{s}\left(\V{r}_1\right) = - k \sqrt{Z}  \sum_{\alpha} \left[ \utbSmat_{4} \M{I}\right]_\alpha \M{u}_{\alpha}^{\left(1\right)}\left(k\V{r}_1\right).
\end{equation}

\section{Transition Matrix for a Spherical Shell}
\label{App:TmatSphere}
There are several ways to determine the transition matrix, the core operator of the T-matrix method. Originally the solution was presented in \cite{Waterman1965} using the Null-field method which represents an efficient method of solving the scattering in terms of surface integrals in the basis of spherical vector waves. Just as it is possible to determine the transition matrix, it is also possible to apply the same procedure to determine other scattering operators.

The transition matrix of a PEC spherical shell forms a diagonal matrix with elements
\begin{equation}
    \label{eq:TPEC}
    T_\alpha = -\dfrac{ \T{R}_\alpha^{\left(1\right)}\left(ka\right)}{ \T{R}_\alpha^{\left(4\right)}\left(ka\right)},
\end{equation}
while the internal scattering matrix $\M{\Gamma}$ contains diagonal elements
\begin{equation}
    \label{eq:GPEC} 
    \Gamma_\alpha = -\dfrac{ \T{R}_\alpha^{\left(4\right)}\left(ka\right)}{ \T{R}_\alpha^{\left(1\right)}\left(ka\right)},
\end{equation}
with~$a$ being the radius of the shell. Notice that all the relations in this section are derived from the equality of tangent components of the fields at the interface and for this reason only $\alpha = \{\tau \sigma m l \}$ for $\tau=1,2$ should be used.

As with~\eqref{eq:TPEC}, the transition matrix of a sphere made of a homogeneous and isotropic dielectric can be determined as
\begin{equation}
    \label{eq:Tsphere}
    T_\alpha =
    - \dfrac{\T{R}_\alpha^{\left(1\right)}\left(ka\right)}{\T{R}_\alpha^{\left(4\right)}\left(ka\right)} \dfrac{1 - \dfrac{Z_\odot}{Z}\dfrac{\T{R}_\alpha^{\left(1\right)}\left(k_\odot a\right)\T{R}_{\bar{\alpha}}^{\left(1\right)}\left(ka\right)}{\T{R}_{\bar{\alpha}}^{\left(1\right)}\left(k_\odot a\right)\T{R}_\alpha^{\left(1\right)}\left(ka\right)} }{1 - \dfrac{Z_\odot}{Z}\dfrac{\T{R}_\alpha^{\left(1\right)}\left(k_\odot a\right)\T{R}_{\bar{\alpha}}^{\left(4\right)}\left(ka\right)}{\T{R}_{\bar{\alpha}}^{\left(1\right)}\left(k_\odot a\right)\T{R}_\alpha^{\left(4\right)}\left(ka\right)}},
\end{equation}
where quantities with the subscript ${}_\odot$ belong to the material of the sphere and $\bar{\alpha} = \bar{\tau}\sigma m l$ is the index dual to index $\alpha$, where $\bar{1} = 2$ and $\bar{2} = 1$.

In addition to the transition matrix of the material sphere, it is also possible to determine matrix~$\M{\Gamma}$ of a spherical cavity inside a homogeneous isotropic dielectric as
\begin{equation}
    \label{eq:Gsphere}
    \Gamma_\alpha = -\dfrac{\T{R}_{\alpha}^{\left(4\right)}\left(k_\odot a\right)}{\T{R}_{\alpha}^{\left(1\right)}\left(k_\odot a\right)}\dfrac{1-\dfrac{Z}{Z_\odot}\dfrac{\T{R}_\alpha^{\left(4\right)}\left(ka\right)\T{R}_{\bar{\alpha}}^{\left(4\right)}\left(k_\odot a\right)}{\T{R}_{\bar{\alpha}}^{\left(4\right)}\left(ka\right)\T{R}_{\alpha}^{\left(4\right)}\left(k_\odot a\right)} }{1-\dfrac{Z}{Z_\odot} \dfrac{\T{R}_\alpha^{\left(4\right)}\left(ka\right)\T{R}_{\bar{\alpha}}^{\left(1\right)}\left(k_\odot a\right)}{\T{R}_{\bar{\alpha}}^{\left(4\right)}\left(ka\right)\T{R}_{\alpha}^{\left(1\right)}\left(k_\odot a\right)} }
\end{equation}
where ${}_\odot$ indicates quantities that pertain to the cavity.
Notice that if the absolute value of the wavenumber in the material of the background medium approaches infinity, then relations~\eqref{eq:Tsphere} and~\eqref{eq:Gsphere} approach those for the \ac{PEC} shell~\eqref{eq:TPEC} and \ac{PEC} cavity~\eqref{eq:GPEC}, respectively.

\section{Relation Between MoM and Spherical Matrices}
\label{App:ZtoT}

Consider a generalized description~\eqref{eq:TmatGeneral} of a scatterer from~Fig.~\ref{fig:TGIllustration}. Consider that this scatterer is also described by matrix~$\M{Z}$. Projector~$\utbSmat_p$ introduced in~Sec.~\ref{sec:Hybrid} can then be used to interlink these two descriptions as 
\begin{equation}
\label{eq:TmatToZ}
\mqty[ \M{T} & \M{\Psi} \\
       \M{\Psi}^\T{T} & \M{\Gamma} ] = 
\mqty[
            -\utbSmat_1\M{Z}^{-1}\utbSmat_1^\T{T} & -\utbSmat_1\M{Z}^{-1}\utbSmat_4^\T{T} + \M{1} \\
            -\utbSmat_4\M{Z}^{-1}\utbSmat_1^\T{T} + \M{1} & -\utbSmat_4\M{Z}^{-1}\utbSmat_4^\T{T}
 ] ,
\end{equation}
which can be deduced from relations between vectors~$\M{a},\M{f}$ and $\M{I},\M{V}$. In this way the spherical wave matrices can be numerically evaluated for an arbitrarily shaped object. Relations~\eqref{eq:TmatToZ} generalize those published in~\cite[(12)]{2013_Kim_TAP} and~\cite[(20)]{2017_Markkanen_JQSRT} by matrices~$\M{\Psi}$ and~$\M{\Psi}^\trans$.

For a volumetric \ac{PEC} obstacle, relation~\eqref{eq:TmatToZ} is also tightly connected to the Null-field method~\cite{Waterman1965,Kristensson_ScatteringBook}. There, relation~$\M{T} \utbSmat_4 = -\utbSmat_1$ is proposed~\cite[(Chap. 9.1)]{Kristensson_ScatteringBook} as an algebraic possibility to obtain matrix~$\M{T}$ (In \cite{Kristensson_ScatteringBook}, matrices~$\utbSmat_1,\utbSmat_4$ are, apart from multiplicative constant, denoted as~$\M{R},\M{Q}$.). For spheroidal objects this method gives acceptable results. For object of complex shape, however, the relation~$\M{T} \utbSmat_4 = -\utbSmat_1$ cannot be precisely inverted, since necessary high order spherical waves make the pseudo-inversion of matrix~$\utbSmat_4$ imprecise. For complex shaped objects, the prescription~\eqref{eq:TmatToZ} is therefore a preferred way to obtain matrix~$\M{T}$. Unlike Null-field method, a matrix~$\M{T}$ can in this way be obtained also for planar objects.

\section{Relation Between Transition Matrix and Characteristic Modes}
\label{app:CM}
Let us assume a lossless scatterer from~Fig.~\ref{fig:TGIllustration}. On its external side, the characteristic mode decomposition is defined by~\eqref{eq:CM1}, which can be generally rewritten as
\begin{equation}
    \label{eq:CM2}
    \M{Z}\M{I}_n = \left(1 + \T{j}\lambda_n \right) \M{R}\M{I}_n
\end{equation}
in which $\M{Z}$ represents the impedance matrix. Multiplying~\eqref{eq:CM2} from the left by~$\M{Z}^{-1}$ and using relation\footnote{This relation is only valid for a lossless scatterer.}~$\M{R} = \utbSmat_1^\T{T} \utbSmat_1$, which was derived in~\cite{TayliEtAl_AccurateAndEfficientEvaluationofCMs}, it is possible to deduce that
\begin{equation}
    \label{eq:CM3}
    \M{I}_n = \left(1 + \T{j}\lambda_n \right) \M{Z}^{-1} \utbSmat_1^\T{T} \utbSmat_1\M{I}_n.
\end{equation}
Further multiplication from the left by projector~$\utbSmat_1$ and the usage of relation~\eqref{eq:TmatToZ} between matrices~$\M{Z}$ and~$\M{T}$ leads to
\begin{equation}
    \label{eq:CM4}
    -\utbSmat_1 \M{I}_n =\left(1 + \T{j}\lambda_n \right) \M{T}  \utbSmat_1\M{I}_n.
\end{equation}
Since~$\M{f}_{1,n} = -\utbSmat_1 \M{I}_n$ is an eigenvector of expansion coefficients corresponding to spherical waves propagating outwards, the characteristic modes might also be understood as eigenvectors of matrix~$\M{T}$
\begin{equation}
   \label{eq:CM5}
   \M{T}  \M{f}_{1,n} = -\dfrac{1}{1 + \T{j}\lambda_n} \M{f}_{1,n},
\end{equation}
which is consistent with~\cite{Garbacz_TCMdissertation}, \cite{HarringtonMautz_TheoryOfCharacteristicModesForConductingBodies}. This relation also shows that characteristic modes of a spherical scatterer are spherical vector waves, since, in that case, matrix~$\M{T}$ is diagonal. As an example, characteristic modes of a PEC spherical shell can be evaluated by combining~\eqref{eq:CM5} and~\eqref{eq:TPEC}, which leads to
\begin{equation}
    \label{eq:CM6}
   \lambda_\alpha = \T{j} \left(1 - \dfrac{ \T{R}_\alpha^{\left(4\right)}\left(kr\right)}{ \T{R}_\alpha^{\left(1\right)}\left(kr\right)} \right).
\end{equation}
Employing~\eqref{eq:AppA:Rfunc}, this relation is identical to~\cite[(16) and (17)]{2017_Losenicky_Comment_Pfeiffer}.

For lossy scatterers, the decomposition~\eqref{eq:CM5} loses its relation to characteristic modes since definition~\eqref{eq:CM1} will not be equivalent to~\eqref{eq:CM2}. It is also worth mentioning that, as opposed to~\eqref{eq:CM1} or~\eqref{eq:CM2}, taking~\eqref{eq:CM5} as the defining relation of characteristic modes\footnote{It seems to be their original definition, see~\cite{1948_Montgomery_Principles_of_Microwave_Circuits} and~\cite{Garbacz_TCMdissertation}.} can solve the long-lasting issues with characteristic modes of dielectric bodies~\cite{YlaOijala_PMCHWTBasedCharacteristicModeFormulationsforMaterialBodies}. This treatment only demands the knowledge of the transition matrix obtained from volume~\cite{HarringtonMautzChang_CharacteristicModesForDielectricAndMagneticBodies} or surface integral equations~\cite{ChangHarrington_AsurfaceFormulationForCharacteristicModesOfMaterialBodies} unifying thus both approaches. The eigenmodes of matrix~$\M{T}$ can also shed new light on the connection between characteristic modes and natural modes~\cite{Huang_StudyontheRelationshipsbetweenEigenmodesNaturalModesandCharacteristicModes}.

\section{Power Balance}
\label{app:PowerBalance}
Within a time harmonic steady state of convention~$\T{exp}\{\T{j}\omega t\}$, where~$\omega$ is the angular frequency, the balance of a cycle mean electromagnetic power is described as~$P_{\V{J}} = P_S + P_\T{lost}$, where
\begin{equation}
\label{eq:appD:Poynting1}
P_{\V{J}} = - \dfrac{1}{2} \T{Re} \int \limits_V \V{E} \cdot \V{J}^*_\T{i} \T{d}V
\end{equation}
is the power supplied by current source~$\V{J}_\T{i}$ in volume~$V$,
\begin{equation}
\label{eq:appD:Poynting2}
P_S = \dfrac{1}{2} \T{Re} \oint \limits_S \left( \V{E} \times \V{H}^* \right) \cdot \T{d}\V{S}
\end{equation}
is the net power passing surface~$S$, and~$P_\T{lost}$ is the power lost in volume~$V$.

Imagine that surface~$S$ is a spherical shell centered at origin and passing solely through material background. For the setup treated in this paper, the only source of power~\eqref{eq:appD:Poynting1} is object~$\varOmega_\T{a}$, which, when being within volume~$V$, generates
\begin{equation}
\label{eq:appD:Poynting1a}
P_{\V{J}} = \dfrac{1}{2} \T{Re} \left\{ \M{I}^\T{H} \M{V}^\T{i} \right\}.
\end{equation}

Assuming that the spherical wave decomposition
\begin{equation}
    \label{eq:Edecomp}
    \V{E} \left(\V{r}\right) = k \sqrt{Z}\sum_\alpha a_{\alpha} \,\UFCN{\alpha}{1}{k\V{r}} + f_{\alpha} \,\UFCN{\alpha}{4}{k\V{r}}
\end{equation}
is known on surface~$S$, the net power passing the surface might be evaluated as~\cite{Kristensson_ScatteringBook}
\begin{equation}
\label{eq:appD:Poynting2a}
P_S = \dfrac{1}{2} \left( \left| \M{f} \right|^2 + \T{Re} \left\{ \M{a}^\T{H} \M{f} \right\} \right),
\end{equation}
where the outward and inward fluxes are naturally separated.

The power lost within object~$\varOmega_\T{a}$ is evaluated as
\begin{equation}
    P_\T{lost}^\T{a} = \dfrac{1}{2} \M{I}^\T{H} \T{Re} \left\{ \M{Z}_\rho  \right\} \M{I}.
\end{equation}
The evaluation of loss in object~$\varOmega_\T{p}$ is most easily approached via~\eqref{eq:appD:Poynting2a}.

With the above knowledge, it might be stated that the total radiated power in the scenario of~Fig.~\ref{fig:external} reads
\begin{equation}
\begin{split}
    P_\T{rad} &= \dfrac{1}{2} \left| \M{f}_1 - \utbSmat_1 \M{I} \right|^2 = \\ 
    &= \dfrac{1}{2} \left| \M{f}_1 \right|^2 +  \dfrac{1}{2} \M{I}^\T{H} \T{Re} \left\{ \M{Z}_0  \right\} \M{I} - \T{Re} \left\{ \M{f}_1^\T{H} \utbSmat_1 \M{I} \right\},
\end{split}
\end{equation}
where the term~$- \utbSmat_1 \M{I}$ represents the outgoing spherical waves generated by object~$\varOmega_\T{a}$, see~\eqref{eq:f2}. The total lost power reads
\begin{equation}
\begin{split}
    P_\T{lost} &= \dfrac{1}{2} \M{I}^\T{H} \T{Re} \left\{ \M{Z}_\rho  \right\} \M{I} \\ 
    &- \dfrac{1}{2} \left( \left| \M{f}_1 \right|^2 + \T{Re} \left\{ \left( \M{a}^\T{i} - \utbSmat_4 \M{I} \right)^\T{H} \M{f}_1 \right\} \right).
\end{split}
\end{equation}

Summing the lost and radiated power gives the total cycle mean power supplied by all sources which reads
\begin{equation}
   P_\T{rad} + P_\T{lost} = P_{\V{J}} - \dfrac{1}{2} \T{Re} \left\{ \M{a}^{\T{i},\T{H}} \left(\M{f}_1 - \utbSmat_1 \M{I} \right) \right\}, 
\end{equation}
where the last term (including the minus sign) is the cycle mean power supplied by spherical waves~$\M{a}^\T{i}$.

The total power supplied by the sources can also be evaluated from~\eqref{eq:ExternalFull} as
\begin{equation}
P_\T{rad} + P_\T{lost} = \dfrac{1}{2} \T{Re} \left\{ \mqty[ \M{I} \\ \M{f}_1 \\ \M{a}_1
        ]^\T{H}
    \mqty[ \utbSmat_1^\T{T} \M{a}^\T{i} + \M{V}^\T{i} \\ -\M{a}^\T{i} \\\M{0} ] \right\}.
\end{equation}

The power balance in the scenarios of Fig.~\ref{fig:internal} and Fig.~\ref{fig:unified} can be obtained analogously.


\section{Solver Settings}
\label{app:Solvers}
This appendix summarizes the setup of commercial solvers that were used for comparison with the proposed method.
\begin{itemize}
\item FEKO (ver. 2021.0.1,~\cite{feko2021}) has been used with mesh structure of the dipole imported using NASTRAN data format~\cite{nastran} and dielectric spherical shell modeled using built-in CAD tools. The mesh structure of the dipole consisted of 396~triangles, and its excitation was realized by a delta-gap feed (mesh port). The spherical shell was discretized into 35390~tetrahedra. Analysis of the antenna was done using \ac{MoM} solver, and the spherical shell was analyzed using \ac{FEM} solver. The cube scenario, the same dipole models were used, and the material cube was described by a mesh structure made of 51800~tetrahedrons.
\item CST (ver. 2021.01,~\cite{cst2021}) has been used with all objects modeled using build-in CAD tools. The dipole was modeled as a strip with a centered physical gap of length~$w/5$. The excitation of this structure was realized using a discrete s-parametric port with terminal impedance~$Z_0 = 50 \, \Omega$. The model was, in this case, discretized into $5.96 \cdot 10^6$ cells, and its analysis was performed using a time-domain solver. In the case of the cube scenario, the same approach as in previous cases have been used, resulting in a model discretized into $1.6 \cdot 10^6$ cells.
\end{itemize}


\begin{IEEEbiography}[{\includegraphics[width=1in,height=1.25in,clip,keepaspectratio]{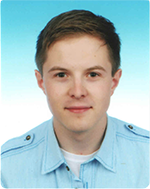}}]{Vit Losenicky}
received the M.Sc. degree in electrical engineering from the Czech Technical University in Prague, Czech Republic, in 2016. 

He is now working towards his Ph.D. degree in the area of electrically small antennas and numerical techniques.
\end{IEEEbiography}

\begin{IEEEbiography}[{\includegraphics[width=1in,height=1.25in,clip,keepaspectratio]{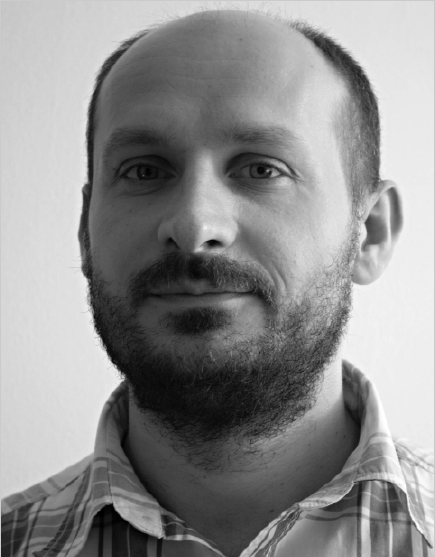}}]{Lukas Jelinek}
received his Ph.D. degree from the Czech Technical University in Prague, Czech Republic, in 2006. In 2015 he was appointed Associate Professor at the Department of Electromagnetic Field at the same university.

His research interests include wave propagation in complex media, general field theory, numerical techniques and optimization.
\end{IEEEbiography}

\begin{IEEEbiography}[{\includegraphics[width=1in,height=1.25in,clip,keepaspectratio]{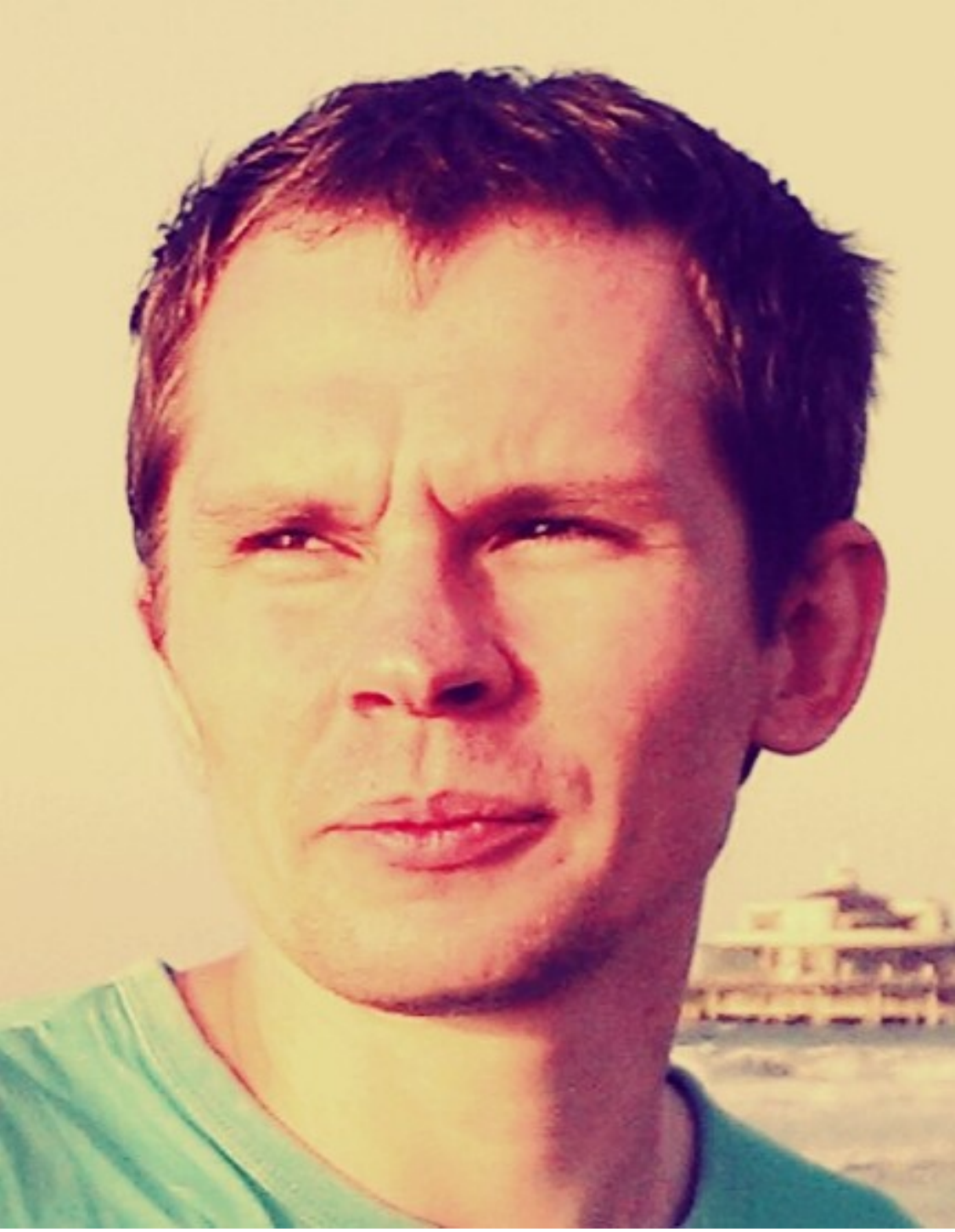}}]{Miloslav Capek}
(M'14, SM'17) received the M.Sc. degree in Electrical Engineering 2009, the Ph.D. degree in 2014, and was appointed Associate Professor in 2017, all from the Czech Technical University in Prague, Czech Republic.
	
He leads the development of the AToM (Antenna Toolbox for MATLAB) package. His research interests are in the area of electromagnetic theory, electrically small antennas, numerical techniques, fractal geometry, and optimization. He authored or co-authored over 100~journal and conference papers.

Dr. Capek is member of Radioengineering Society and Associate Editor of IET Microwaves, Antennas \& Propagation.
\end{IEEEbiography}

\begin{IEEEbiography}[{\includegraphics[width=1in,height=1.25in,clip,keepaspectratio]{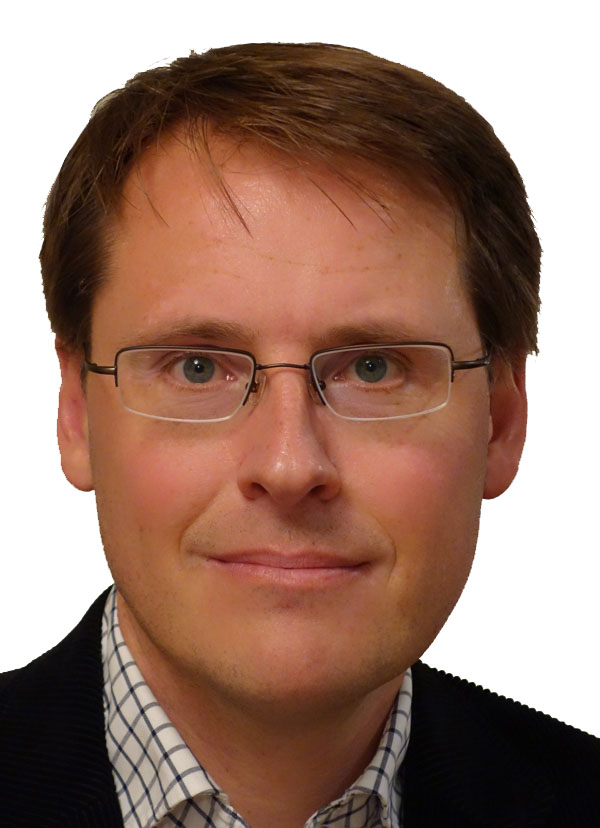}}]{Mats Gustafsson}
received the M.Sc. degree in Engineering Physics 1994, the Ph.D. degree in Electromagnetic Theory 2000, was appointed Docent 2005, and Professor of Electromagnetic Theory 2011, all from Lund University, Sweden.

He co-founded the company Phase holographic imaging AB in 2004. His research interests are in scattering and antenna theory and inverse scattering and imaging. He has written over 100 peer reviewed journal papers and over 100 conference papers. Prof. Gustafsson received the IEEE Schelkunoff Transactions Prize Paper Award 2010, the IEEE Uslenghi Letters Prize Paper Award 2019, and best paper awards at EuCAP 2007 and 2013. He served as an IEEE AP-S Distinguished Lecturer for 2013-15.
\end{IEEEbiography}

\end{document}